\documentclass[11pt,a4paper]{article}
\usepackage{amsmath,amssymb}
\usepackage{epsfig,graphicx}
\usepackage[section]{placeins}
\begin{document}
\title{\vskip -70pt
\begin{flushright}
{\normalsize DAMTP-2009-26}
\\{\normalsize DCPT-09/25}
\end{flushright}
\vskip 50pt
\bf Light Nuclei of Even Mass Number\\ in the Skyrme Model
\author{
R.A. Battye$^{1}$\footnote{Richard.Battye@manchester.ac.uk}\,, N.S. Manton$^{2}$\footnote{N.S.Manton@damtp.cam.ac.uk}\,, P.M. Sutcliffe$^{3}$\footnote{P.M.Sutcliffe@durham.ac.uk} \,\,\,and
\,\,S.W. Wood
$^{2}$\footnote{S.W.Wood@damtp.cam.ac.uk}\\
\\$^{1}$ {\sl{Jodrell Bank Centre for Astrophysics,}}
\\{\sl{University of Manchester, Manchester M13 9PL, U.K.}}
\\
\\$^{2}$ {\sl{Department of Applied Mathematics and Theoretical Physics,}}
\\{\sl{University of Cambridge, Wilberforce Road, Cambridge CB3 0WA, U.K.}}
\\
\\$^{3}$ {\sl{Department of Mathematical Sciences,}}
\\{\sl{Durham University, South Road, Durham DH1 3LE, U.K.}}
}
}
\maketitle
\begin{abstract}
We consider the semiclassical rigid-body quantization of Skyrmion solutions
of mass numbers $B=4$, 6, 8, 10 and 12. We determine the allowed quantum states for each Skyrmion, and find 
that they often match the observed states of nuclei.
The spin and isospin inertia tensors of these Skyrmions are 
accurately calculated for the first time, and are used to determine the excitation energies
of the quantum states.
We calculate the energy level splittings, using a suitably chosen parameter set for each mass number.
We find good qualitative and encouraging quantitative agreement with experiment.
In particular, the rotational bands of beryllium-8 and carbon-12, along with isospin 1 triplets and isospin 2 quintets, 
are especially well reproduced. We also predict the existence of states that have not yet been observed, 
and make predictions for the unknown quantum numbers of some observed states.

\end{abstract}
\section{Introduction}
The SU(2) Skyrme model provides a geometrical picture of nuclear physics in which nuclei are identified with 
the topological soliton solutions of the model, known as Skyrmions \cite{skyrme}. The model has several advantages over
conventional nuclear models. Firstly, single Skyrmions, which are identified with nucleons, are
found to merge to some extent and to lose their individual identities in the solutions describing larger nuclei. 
This captures an important feature of nuclei with individual nucleons close together, and is something that
can not be achieved in conventional point nucleon models. Secondly, the Skyrme
Lagrangian is defined in terms of only three parameters. For each mass number $B$ we fit these parameters to the mass
and size of the nucleus of zero isospin that has mass number $B$, to obtain predictions with reasonable quantitative accuracy. This is unlike potential and 
shell model calculations which require many finely tuned parameters. 

The model is a pion field theory, and is an extension of the nonlinear sigma model. The pion fields
$\boldsymbol{\pi}(x)$ are combined into an SU(2)-valued scalar field,
the Skyrme field
\begin{equation}
U(x)=\sigma (x) 1_2 +i\boldsymbol{\pi} (x)\cdot \boldsymbol{\tau}\,,
\end{equation}
where $\boldsymbol{\tau}$ denotes the triplet of Pauli matrices and 
$\sigma^2 + \boldsymbol{\pi}\cdot \boldsymbol{\pi}=1$. The Lagrangian density is \cite{mantonbook}
\begin{equation}
\mathcal{L} = \frac{F_\pi^2}{16}\,\hbox{Tr}\,\partial_\mu U
\partial^\mu U^{\dag} + \frac{1}{32e^2}\,\hbox{Tr}\,[\partial_\mu U U^{\dag},
\partial_\nu U U^{\dag}][\partial^\mu U U^{\dag},
\partial^\nu U U^{\dag}] + \frac{1}{8} m_\pi ^2 F_\pi^2\,\hbox{Tr}\,(U-1_2) \,,
\end{equation}
where $F_\pi$ is the pion 
decay constant, $e$ is a dimensionless parameter and $m_\pi$ is the pion mass.
Using energy and length units of $F_\pi / 4e$ and
$2/eF_\pi$ respectively, the Skyrme Lagrangian can be rewritten as
\begin{equation} \label{eq:l}
L=\int \left\{ -\frac{1}{2}\,\hbox{Tr}\,(R_\mu R^\mu) 
+ \frac{1}{16}\,\hbox{Tr}\,([R_\mu,R_\nu][R^\mu,R^\nu]) 
+ m^2\,\hbox{Tr}\,(U - 1_2) \right\} d^3 x \,,
\end{equation}
where $R_\mu
= \partial_\mu UU^{\dag}$, and the dimensionless pion mass
$m=2m_\pi /eF_\pi$. As usual, the Lagrangian splits
into kinetic and potential parts as $L = T-V$, with $T$ quadratic in
the time derivative of the Skyrme field. The potential energy $V$ is given by
\begin{equation}
V=\int\left\{-\frac{1}{2}\,\hbox{Tr}\,(R_iR_i)-
\frac{1}{16}\,\hbox{Tr}\,([R_i,R_j][R_i,R_j])-\,
m^2\hbox{Tr}(U-1_2)\right\}d^3x\,.
\end{equation}
The Skyrmions are minima of the potential energy and are static. 
They are labelled by their topological charge, $B$,
the degree of the map $U:{\mathbb{R}}^3 \rightarrow \rm{SU(2)}$, which is well-defined
provided $U \rightarrow 1_2$ at spatial infinity, and is given by the integral 
\begin{equation} 
B = \int B_0(x) \, d^3 x\,,
\end{equation} 
where 
\begin{equation}
\label{eq:cur}B_\mu(x)=\frac{1}{24\pi^2}\,
\epsilon_{\mu\nu\alpha\beta}\,\hbox{Tr}\, \left(\partial^\nu
UU^{\dag}\partial^\alpha U U^{\dag}\partial^\beta U U^{\dag}\right) \,.
\end{equation}
We denote the minimised potential energy by ${\cal{M}}_B$.
One interprets a charge $B$ Skyrmion, after quantization,
as a nucleus of mass number $B$. In this picture, nucleons and nuclei
arise purely from the pion field and no explicit nucleonic sources are needed.

It is necessary to semiclassically quantize the Skyrmions as rigid
bodies. The Skyrme Lagrangian is invariant under
rotations in space. It is also invariant under the transformations 
$U \rightarrow AUA^\dag$, where $A \in \rm{SU(2)}$. This is isospin symmetry. 
The rotational and isorotational degrees of freedom are treated as collective
coordinates and the Skyrmions are quantized in their rest frame by canonical methods.
In this way the Skyrmions acquire spin and isospin. An advantage of this model over other
nuclear models that involve collective rotational motion
is that it incorporates isospin excitations. In the Skyrme
model, the vacuum solution $U=1_2$ is isospin invariant, but for each
classical Skyrmion solution, isospin symmetry as well as rotational
symmetry is spontaneously broken. These symmetries are restored by the
collective coordinate quantization.

The kinetic energy of a rigidly rotating Skyrmion (ignoring the rather
trivial translational motion) is of the form
\begin{equation} \label{eq:t}
T=\frac{1}{2}a_i U_{ij} a_j - a_i W_{ij} b_j + \frac{1}{2}b_i V_{ij}b_j \,, 
\end{equation}
where $b_i$ and $a_i$ are the angular velocities in space and isospace 
respectively, and $U_{ij}$, $V_{ij}$ and $W_{ij}$ are inertia tensors \cite{bc,mmw}.
The momenta conjugate to $b_i$ and $a_i$ are the body-fixed spin and isospin operators
$L_i$ and $K_i$. The quantum Hamiltonian $H$ is obtained by re-expressing $T$ in terms 
of these operators. The space-fixed spin and isospin operators are denoted $J_i$ and $I_i$
respectively. Note that ${\mathbf{J}}^2={\mathbf{L}}^2$ and ${\mathbf{I}}^2={\mathbf{K}}^2$.

Finkelstein and Rubinstein showed that it is possible to quantize a $B=1$ Skyrmion
as a fermion, and showed that for even (odd) $B$ the spin and isospin are integers (half-integers) \cite{fr}.
Discrete symmetries of the classical Skyrmion solutions\footnote{Only
  the $B=1$ and $B=2$ Skyrmions possess continuous symmetries.} give rise to further Finkelstein--Rubinstein
(FR) constraints on the space of quantum states $|\Psi \rangle$. These constraints are of the form
\begin{equation}
e^{i\theta_2 {\mathbf{n}}_2 \cdot {\mathbf{L}}}e^{i\theta_1 {\mathbf{n}}_1 \cdot {\mathbf{K}}}|\Psi
\rangle = \chi_{\rm{FR}} |\Psi \rangle\,,
\end{equation}
where ${\mathbf{n}}_1$, ${\mathbf{n}}_2$ and $\theta_1$, $\theta_2$
are, respectively, the axes and angles defining the rotations in
isospace and space associated with a particular symmetry, and
$\chi_{\rm{FR}} = \pm 1$. The FR signs, $\chi_{\rm{FR}}$, define a one-dimensional
representation of the symmetry group of the Skyrmion. Krusch has found
a convenient way to calculate them \cite{krusch}, and we use this
method here. A basis for the wave functions $|\Psi \rangle$ is
given by the products $|J,L_3 \rangle \otimes |I,K_3 \rangle$,
the tensor products of states for a rigid body in space and a
rigid body in isospace. $J$ and $I$ are the total spin and isospin
quantum numbers, $L_3$ and $K_3$ the projections on to the third
body-fixed axes, and the space projection labels (which are the
physical third components of spin and isospin) are suppressed. The
FR constraints only allow a subset of these products as physical states. A 
parity operator is introduced by considering a Skyrmion's reflection 
symmetries. Quantum states are therefore labelled by the usual
quantum numbers: spin-parity $J^{\pi}$ and isospin $I$.

The inclusion of the third term in the Lagrangian density, which involves the pion mass, has
a significant effect on the shapes and symmetries of the classical Skyrmion solutions. This effect is more
marked for larger values of $B$. For zero pion mass, the Skyrmions 
with $B$ up to 22 and beyond resemble hollow polyhedra, with their baryon density concentrated in
a shell of roughly constant thickness, surrounding a region in which the baryon density is very
small \cite{battyesym}. This disagrees with the approximately uniform baryon density observed in the interior of
real nuclei. Fortunately, it has been established that the hollow polyhedral solutions for
$B \geq 8$ do not remain stable when the pion mass is set at a physically
reasonable value, with $m \approx 1$ \cite{battyemas}. 
This is because in
the interior of the hollow polyhedra the Skyrme field is very close to $U=-1_2$, and here the
pion mass term gives the field a maximal potential energy, and hence instability. This instability results in the
interior region splitting into separate smaller subregions. The stable Skyrmion solutions are
found to exhibit clustering: small Skyrmion solutions, such as the cubically symmetric
$B=4$ solution, appear as substructures within larger solutions \cite{bms}. This is a very encouraging
development as it has been believed for some time that alpha-particles exist 
as stable substructures inside heavier nuclei. The most remarkable success of the alpha-particle
model is in its prediction for the binding energies of nuclei which can be formed out of an
integral number of alpha-particles. 

In earlier work, the $B=4$, 6 and 8 Skyrmions have been
quantized \cite{mw,mmw}, using some approximations to the Skyrmion solutions based on the rational map ansatz \cite{houghton}. 
The
inertia tensors used had the right symmetries, but not the correct
numerical values. In this paper we consider the $B=4$, 6, 8, 10 and 12
Skyrmions, using a consistent numerical scheme to recalculate all
these solutions. Each of these Skyrmions can be viewed as being built
up from $B=4$ cubes (it is possible to regard the $B=6$ Skyrmion as being
made up of a $B=4$ cube and a $B=2$ torus, and the $B=10$ solution as consisting of
two $B=4$ cubes together with two $B=1$ Skyrmions).
We numerically relax field configurations to the stable Skyrmions, 
for various values of the dimensionless pion mass
$m$, and then compute their static energies ${\cal{M}}_B$, charge 
radii $\left\langle r^2 \right\rangle ^{1/2}$, and
inertia tensors. Appendix A tabulates the calculated numerical values 
for $m=0.5$, 1 and 1.5, and describes the quadratic interpolation 
method used to estimate these quantities for any given $m$.
In the next section we describe our method of choosing $m$ and of calibrating the model 
to ensure that it provides quantitatively accurate predictions of nuclear properties.

For each Skyrmion we determine all the FR-allowed quantum states and their excitation energies, 
working up to spin and isospin values just beyond the edge of what is experimentally accessible.
Our rigid-body quantization leads to an infinite tower of quantum states (the $B=1$ Skyrmion, for example, has quantum states for all half-integer values of spin). 
However, in practice we expect a Skyrmion to deform as it spins (this is known
to occur for the $B=1$ Skyrmion \cite{spinning}), and this disallows many higher spin states.
We are therefore not concerned that in some cases we predict higher spin states that are
not experimentally observed. It is also possible that a Skyrmion might break up as it spins, a phenomenon that is known to occur for real nuclei.

\section{Calibration}
The model was first calibrated by Adkins, Nappi and Witten, by 
fitting the model in the $B=1$ sector to the masses of the proton 
and delta resonance \cite{anw,an}. In a recent paper a new calibration was obtained by 
equating the mass and size of the quantized $B=6$ Skyrmion to the mass and size of the lithium-6 nucleus \cite{mw}. However this was performed using the 
approximate Skyrmion found using the rational
map ansatz. Evidently, there are many possible ways in which the model may be calibrated.

The three parameters of the model are the pion decay constant $F_\pi$ (experimentally $186\,\hbox{MeV}$), the pion mass $m_\pi$ (experimentally $138\,\hbox{MeV}$), and 
the dimensionless parameter $e$. Strictly speaking, one may argue that we are only free to set $e$, as the other constants are fixed by experiment. However,
we consider it permissible to vary $F_\pi$, as we consider it to be a renormalised pion decay constant. We also allow $e$ to vary with mass number,
$e=e(B)$, in order to get better agreement with experiment. The length and classical energy scales are $2/eF_\pi$ and  $F_\pi /4e$, respectively. 
We recall that the quantum Hamiltonian for a rigidly rotating body is equal to the squared angular momentum operator divided by twice the moment of inertia of the body
\cite{landau}.
The moment of inertia has units equal to the mass scale multiplied by the square of the length scale:
$(F_\pi /4e)\times (2/eF_\pi)^2 = 1/e^3F_\pi$. The quantum energy scale is its reciprocal, $e^3 F_\pi$. The total energy of a quantum state of a Skyrmion is therefore equal to
$(F_\pi /4e){\mathcal{M}}_B+e^3F_\pi E_Q$, where $E_Q$ is the quantum kinetic energy of the state.
The ratio of the total energy of a quantum state to the classical energy of the Skyrmion is $1+4e^4 (E_Q/{\mathcal{M}}_B)$. The ratios of the quantum energies of Skyrmion states are therefore sensitive to the value of $e$.

The classical Skyrmion solutions match the experimental pion tails of nuclei if we use the physical value of $m_\pi$ of 138\,MeV. For this reason, we keep $m_\pi$ fixed at this value. For each $B$, we choose $m$ (and therefore fix the length scale, as $m=2m_\pi /eF_\pi$) such that 
the calculated mean charge radius agrees with that of the nucleus with zero isospin with this value of $B$. Within the Skyrme model, the 
mean charge radius of a nucleus with 
zero isospin is estimated to be the square root of 
\begin{equation} 
\langle r^2 \rangle = \frac{\int r^2\,B_0(x) \, d^3 x}{\int B_0(x) \, d^3 x}\,,
\end{equation}
since the electric charge density is half the baryon density \cite{mw}.
We then set the classical energy scale such that 
the Skyrmion mass agrees with the nuclear mass, which is done by setting 
$F_\pi /4e$ to be the nuclear mass divided by ${\cal{M}}_B$. We use
the actual nuclear masses, not taking into account small quantum
energies which are of the order of 0.1\% of the nuclear mass\footnote{
Lithium-6 and boron-10 have quantum energies associated with the non-zero spins of their ground states ($1^+$ and $3^+$ respectively).
We calculate the energy of the $1^+$ state of the $B=6$ Skyrmion to be 1.6\,MeV (see section 4), which is negligible compared to the mass of lithium-6, 5601\,MeV.
An estimate of the boron-10 mass with its quantum energy removed can be obtained by taking the average of the masses of beryllium-10 and carbon-10 and 
subtracting from this our estimate of the $0^+$ state with isospin 1, 10.4\,MeV (see section 6). This yields a value of 9319\,MeV, which is very close to the 
accepted mass of boron-10, 9327\,MeV.}.
It is
found that in each case $m$ takes a value 
between 0.6 and 1.2. The experimental data, the values of $m$ used, and the length and energy scales are listed in Table 1\footnote{The mean charge radius of beryllium-8 has not been measured, due to its instability. Here we use the value for its isobar lithium-8.}.

\begin{table}[ht]
\centering
\begin{tabular}{|l|l|l|l|l|l|l|l|}
\hline
$B$ & Nucleus & Mean charge & Mass & $m$ & Length & Classical & Quantum \\
& &  radius & (MeV) & & scale & energy scale & energy scale \\
& & (fm) & & & $2/eF_\pi$ (fm) & $F_\pi/4e$ (MeV) & $e^3F_\pi$ (MeV)\\
\hline
4 & $^4$He & 1.71 & 3727 & 0.820 & 1.173 & 6.168 & 4589\\
6 & $^6$Li & 2.55 & 5601 & 1.153 & 1.648 & 5.753 & 2491\\
8 & $^8$Be & 2.51 & 7455 & 0.832 & 1.190 & 6.336 & 4339\\
10 & $^{10}$B & 2.58 & 9327 & 0.830 & 1.187 & 6.354 & 4350\\
12 & $^{12}$C & 2.46 & 11178 & 0.685 & 0.980 & 6.527 & 6214\\
\hline
\end{tabular}
\caption{Experimental data and calibration for each $B$.}
\end{table} 
A larger length scale, for example in the case of $B=6$, takes into account
loose vibrational motion, and leads to larger moments of inertia, which is desirable. The small length scale for $B=12$ takes into account the compact size of the carbon-12
nucleus. Another reason to use separate parameter sets for each $B$ comes from comparing the rotational bands of beryllium-8 and carbon-12. These nuclei have $0^+$ ground states 
and $2^+$ and $4^+$
excited states, at 3.0\,MeV and 11.4\,MeV respectively for beryllium-8, and at 4.4\,MeV and 14.4\,MeV respectively for carbon-12. 
Naively, one might expect carbon-12 to be larger and heavier than
beryllium-8 (in dimensionless units, the $B=12$ Skyrmion is larger in size and has a larger classical mass than the $B=8$ Skyrmion), 
and for it therefore to have a larger moment of inertia. As the moment of inertia appears in the denominator of the quantum Hamiltonian, this would lead to carbon-12 having a 
smaller rotational band splitting than that of beryllium-8, which is not the case. Evidently, the only way to deal with this problem in our model is to use a different 
parameter set for each. Experimentally, the mean charge radii of the nuclei we are considering are approximately constant for $6 \leq B \leq 12$, whereas the dimensionless Skyrmion 
mean charge radius increases with $B$. The optimised parameter sets allow us to keep the nuclear mean charge radii approximately constant.

\section{$B=4$}
The minimal-energy $B=4$ Skyrmion has octahedral symmetry and a cubic shape. A
surface of constant baryon density is presented in Fig. 1. 
The colour scheme represents the direction in isospace of the associated pion fields.
For regions in space where at least one of the pion fields does not vanish, the
normalised pion field
$\hat{\boldsymbol{\pi}}$ can be defined, and takes values in the unit sphere. We colour this sphere by making
a region close to the north pole white and a region close to the south pole black. On an equatorial band, where
$\hat{\pi}_3$ is small, we divide the sphere into three segments and colour these red, blue and green.
In Fig. 1, opposite faces share the same colour and vertices alternate between black and white. 

For a derivation
of the Skyrmion's quantum states we refer the reader to Ref. \cite{mmw}, and here we only state the main
results. For an earlier discussion, see Ref. \cite{walhout}. The rotational symmetry group of the Skyrmion, $O_h$, is one of the largest point
symmetry groups. This leads to particularly restrictive FR constraints on the space of
allowed states. The $O_h$ symmetry
implies that the inertia tensors are diagonal with $U_{11}=U_{22}$ and
$U_{33}$ different, $V_{ij}$ proportional to the identity matrix and
$W_{ij}=0$. The quantum collective coordinate Hamiltonian is therefore
the sum of a spherical top in space and a symmetric top in isospace,
\begin{equation}
H = \frac{1}{2V_{33}}\mathbf{J}^2 + \frac{1}{2U_{11}}\mathbf{I}^2 +
\left(\frac{1}{2U_{33}}-\frac{1}{2U_{11}}\right)K_3^2\,.
\end{equation}
The inertia tensors are given in Appendix A.1.
The lowest state is a $0^+$ state with isospin 0, which has the quantum numbers of the
alpha-particle in its ground state. The first excited state with
isospin 0 is a $4^+$ state, which has not been experimentally
observed, because of its high energy. The lowest state with 
isospin 1 is a $2^-$ state, which matches the observed isotriplet including the 
hydrogen-4 and lithium-4 ground states. The energy levels are given
in Table 2. Only the quantized kinetic
contributions to the total energies are listed; the
static Skyrmion mass must be added to give the
total energy. For example, the ground state has zero kinetic energy 
and therefore its total energy is precisely 3727\,MeV, as given in Table 1. 
The $4^+$ state has additional kinetic energy of 
$10/V_{33}=129.7 \times 10^{-4}$ in dimensionless units. In the final column of Table 2 we list the values in physical units,
using the $B=4$ conversion factor $e^3F_{\pi}=4589\,$MeV given in Table 1. We overpredict the excitation energy of the $2^-$ isotriplet as 43.2\,MeV compared with an average
experimental value of 23.7\,MeV \cite{tilley4}.

\emph{In summary:} The ground state of helium-4 and the isotriplet of $2^-$ states arise as quantum states of the $B=4$ Skyrmion. The cubic shape of the Skyrmion may become deformed as it spins. It may be for this reason that the $4^+$ state is not observed experimentally.

\begin{figure}[h!]
\begin{center}
\includegraphics[width=3cm]{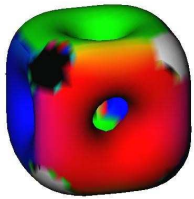}
\caption{A surface of constant baryon density for the $B=4$ Skyrmion. Different
colours indicate different directions of the pion fields.}
\end{center}
\end{figure}

\begin{table}[ht]
\centering
\begin{tabular}{|l|l|l|l|l|}
\hline
$I$ & $J^\pi$ & $E$ ($\times 10^{-4}$) ~~~& $E$ (MeV) \\
\hline
0 & $0^+$ &  0.0 & 0.0 \\
  & $4^+$ & 129.7 & 59.5 \\
\hline
1 & $2^-$ & 94.1 & 43.2 \\
\hline
\end{tabular}
\caption{Energy levels of the quantized $B=4$ Skyrmion.}
\end{table} 

\section{$B=6$}
The $B=6$ Skyrmion has $D_{4d}$ symmetry (see Fig. 2). We refer the reader to Ref. \cite{mmw} for a discussion of its quantization.
The quantum Hamiltonian is that of a system of coupled symmetric
tops:
\begin{equation}
H=\frac{1}{2V_{11}} \mathbf{J}^2
+\frac{1}{2U_{11}} \mathbf{I}^2
+\left(\frac{U_{33}}{2\Delta_{33}}-\frac{1}{2V_{11}}\right)L_3^2
+\left(\frac{V_{33}}{2\Delta_{33}}-\frac{1}{2U_{11}}\right)K_3^2
+\frac{W_{33}}{\Delta_{33}}L_3K_3\,,
\end{equation}
where $\Delta_{33} = U_{33}V_{33}-W_{33}^2$. Its allowed quantum
states are listed in Table 3, together with their energy levels computed using 
the inertia tensors in Appendix A.2. 

\begin{figure}[h!]
\begin{center}
\includegraphics[width=7cm]{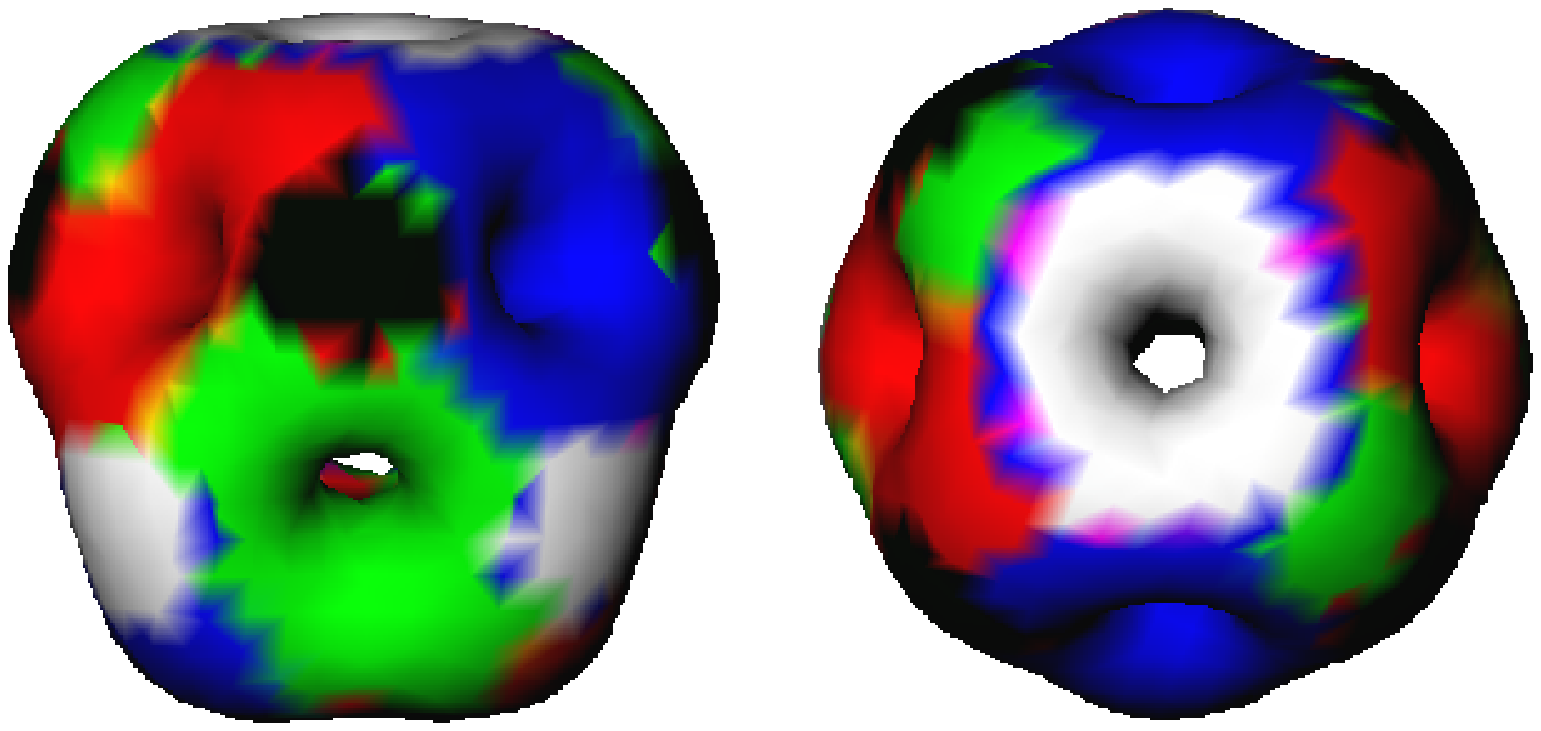}
\caption{A surface of constant baryon density for the $B=6$ Skyrmion (two viewpoints).}
\end{center}
\end{figure}

\begin{table}[ht]
\centering
\begin{tabular}{|l|l|l|l|l|l|}
\hline
$I$ & $J^\pi$ & Wave function ~~~& $E$ ($\times 10^{-4}$) ~~~& $E$ (MeV) \\
\hline
0 & $1^+$ & $|1,0\rangle \otimes |0,0\rangle$ & 6.5 & 1.6 \\
  & $3^+$ & $|3,0\rangle \otimes |0,0\rangle$ & 38.9 & 9.7 \\
  & $4^-$ & $(|4,4\rangle-|4,-4\rangle) \otimes |0,0\rangle$ & 73.5 & 18.3 \\
  & $5^+$ & $|5,0\rangle \otimes |0,0\rangle$ & 97.3 & 24.2 \\
  & $5^-$ & $(|5,4\rangle+|5,-4\rangle) \otimes |0,0\rangle$ & 105.9 & 26.4 \\
\hline
1 & $0^+$ & $|0,0\rangle \otimes |1,0\rangle$ & 47.4 & 11.8 \\
  & $2^+$ & $|2,2\rangle \otimes |1,1\rangle + |2,-2\rangle \otimes |1,-1\rangle$ & 62.6 & 15.6 \\
  &       & $|2,0\rangle \otimes |1,0\rangle$ & 66.8 & 16.6 \\
  & $2^-$ & $|2,2\rangle \otimes |1,-1\rangle + |2,-2\rangle \otimes |1,1\rangle$ & 73.1 & 18.2 \\
  & $3^+$ & $|3,2\rangle \otimes |1,1\rangle - |3,-2\rangle \otimes |1,-1\rangle$ & 82.0 & 20.4 \\
  & $3^-$ & $|3,2\rangle \otimes |1,-1\rangle - |3,-2\rangle \otimes |1,1\rangle$ & 92.5 & 23.1 \\
  & $4^+$ & $|4,2\rangle \otimes |1,1\rangle + |4,-2\rangle \otimes |1,-1\rangle$ & 108.0 & 26.9 \\
  &       & $|4,0\rangle \otimes |1,0\rangle$ & 112.2 & 28.0 \\
  & $4^-$ & $|4,2\rangle \otimes |1,-1\rangle + |4,-2\rangle \otimes |1,1\rangle$& 118.5 & 29.5 \\
  &       & $(|4,4\rangle+|4,-4\rangle) \otimes |1,0\rangle$ & 120.9 & 30.1 \\
\hline
2 & $0^-$ & $|0,0\rangle \otimes (|2,2\rangle - |2,-2\rangle )$ & 137.4 & 34.2 \\
  & $1^+$ & $|1,0\rangle \otimes |2,0\rangle$ & 148.6 & 37.0 \\
  & $1^-$ & $|1,0\rangle \otimes (|2,2\rangle + |2,-2\rangle )$ & 143.9 & 35.8 \\
  & $2^+$ & $|2,2\rangle \otimes |2,1\rangle - |2,-2\rangle \otimes |2,-1\rangle$ & 157.3 & 39.2 \\
  & $2^-$ & $|2,0\rangle \otimes (|2,2\rangle - |2,-2\rangle )$ & 156.9 & 39.1 \\
  &       & $|2,2\rangle \otimes |2,-1\rangle - |2,-2\rangle \otimes |2,1\rangle$ & 167.8 & 41.8 \\
\hline
\end{tabular}
\caption{Energy levels of the quantized $B=6$ Skyrmion.}
\end{table}

\begin{figure}[h!]
\begin{center}
\includegraphics[width=11cm]{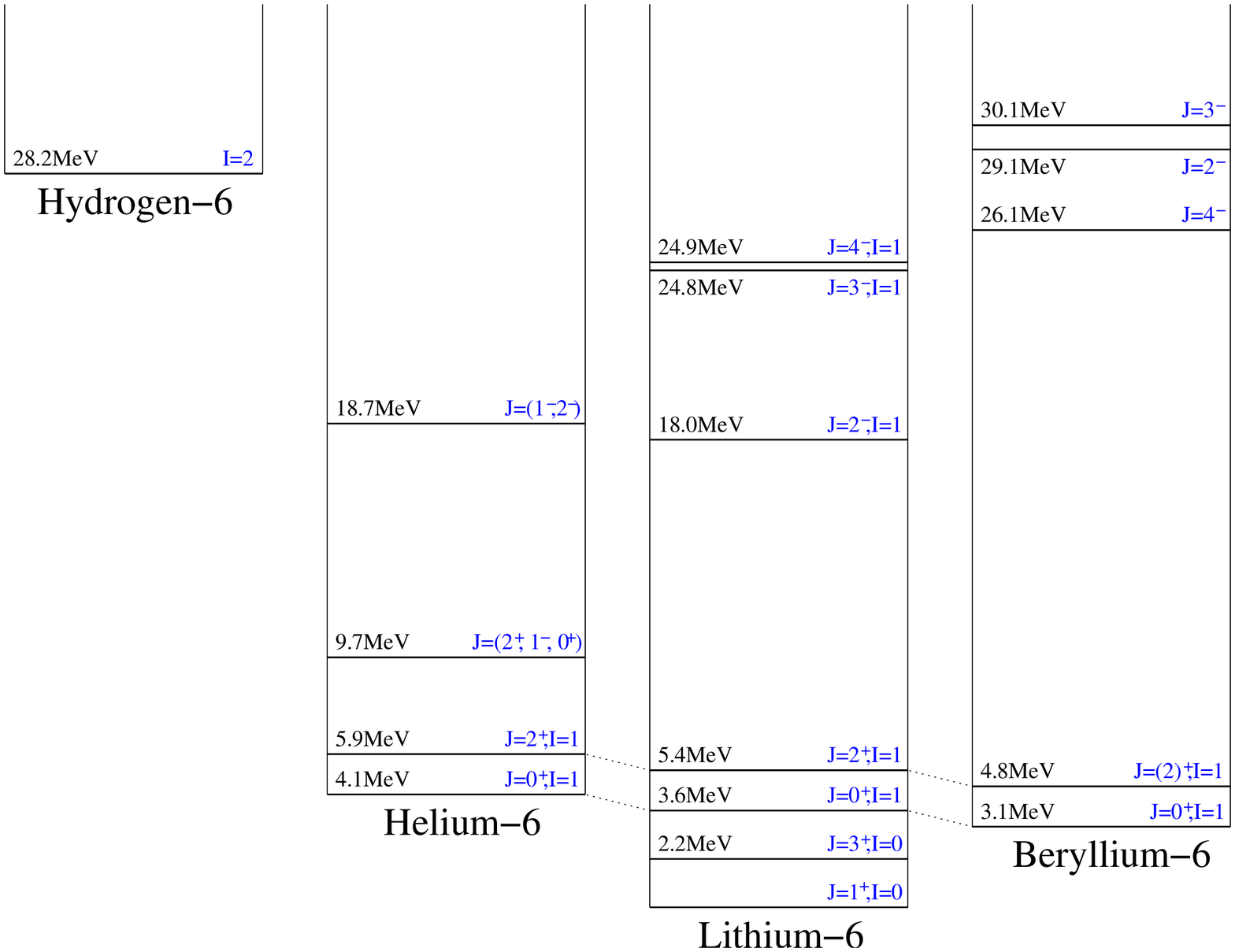}
\caption{Energy level diagram for nuclei of mass number 6. Mass splittings between nuclei are adjusted to eliminate the proton/neutron mass
difference and remove Coulomb effects, as described in Ref. \cite{tilley567}.}
\end{center}
\end{figure}

The Skyrme model qualitatively reproduces the experimental spectrum of
lithium-6 and its isobars, and predicts some further states, including $J^{\pi}=4^-$, $5^+$
and $5^-$ excited states of
lithium-6 with isospin 0. The ground state of lithium-6 is correctly predicted to be a $1^+$ state. We also find a $3^+$ excited state, which is seen experimentally. However
we overpredict its excitation energy by roughly a factor of five. The model does not account for centrifugal stretching nor the 
allowed decay of the $3^+$ state to an alpha-particle plus a deuteron,
which may be the reason for our overprediction.
The lowest allowed state with isospin 1 is a $0^+$ state, which is seen experimentally as an isotriplet
which includes the helium-6 and beryllium-6 ground states. An excited $2^+$ state of this isotriplet exists, which we also find within our model. However again we overpredict its
excitation energy. We predict an additional $2^+$ state with isospin 1.
The lowest allowed negative parity state with isospin 1 is predicted to be a $2^-$ state with excitation energy 18.2\,MeV. We therefore suggest that the observed 9.7\,MeV state of helium-6 is our second $2^+$ state.
Lithium-6 has a $2^-$ state with isospin 1 at 18.0\,MeV. We predict that the 18.7\,MeV state of helium-6 has $J^\pi = 2^-$, and is one of its isotriplet partners. A $2^-$ state of beryllium-6 is observed at 29.1\,MeV. Perhaps this state completes the isotriplet, but its high
energy makes this unclear.
States of lithium-6 and beryllium-6 with $J^\pi = 3^-$ and $4^-$ and with isospin 1 have been experimentally observed. We predict these
states, with roughly the correct excitation energies. We also predict the existence of $3^+$ and $4^+$ states with isospin 1, which have not been seen experimentally.
The ground state of hydrogen-6 has isospin 2, at 28.2\,MeV above the
lithium-6 ground state, and an undetermined spin. The Skyrme model gives the lowest allowed
state with isospin 2 as a $0^-$ state, and its excitation energy to be
34.2\,MeV. Higher spin excited isospin 2 states are also allowed in
the model, but they have not been experimentally observed.

\emph{In summary:} The quantum numbers of the low-lying states of the $B=6$ Skyrmion agree with those of the nuclei of mass number 6, although we overpredict their excitation energies. 
Our calculated values for the excitation energies of the isospin 1 states are, however, quantitatively good.
We have also made predictions for the spins of two excited states of helium-6, and predict that the hydrogen-6 ground state is a $0^-$ state.

\section{$B=8$}
When $m=1$, the stable $B=8$ Skyrmion is $D_{4h}$-symmetric, and
resembles two touching $B=4$ cubes, matching the alpha-particle model picture of
beryllium-8 as an almost bound configuration of two alpha-particles.
A surface of constant baryon density is displayed in Fig. 4.
In Ref. \cite{mmw} we approximated the
$B=8$ Skyrmion as a ``double cube'' of two $B=4$ Skyrmions, 
and made estimates for its inertia tensors, enabling us to estimate its energy levels. 
We refer the reader to this paper for a discussion of its 
quantization. The quantum Hamiltonian is the sum of a symmetric top in space and an asymmetric
top in isospace:
\begin{equation}
H =\frac{1}{2V_{11}}\mathbf{J}^2 
+ \left(\frac{1}{2V_{33}}-\frac{1}{2V_{11}}\right)L_3^2
+\frac{1}{2U_{11}}K_1^2
+\frac{1}{2U_{22}}K_2^2
+\frac{1}{2U_{33}}K_3^2\,.
\end{equation}
The exact inertia tensors are given in Appendix A.3. As anticipated
on symmetry grounds, the inertia tensor $U_{ij}$ has three distinct
eigenvalues (the earlier double cube estimate made two of them equal). 
The energy levels calculated using the exact inertia tensors are listed in Table 4.
\begin{figure}[h!]
\begin{center}
\includegraphics[width=5.2cm]{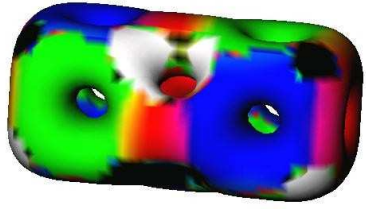}
\caption{A surface of constant baryon density for the $B=8$ Skyrmion, resembling two
$B=4$ cubes.}
\end{center}
\end{figure}

\begin{table}[ht]
\centering
\begin{tabular}{|l|l|l|l|l|}
\hline
$I$ & $J^\pi$ & $n$ & $E$ ($\times 10^{-4}$) ~~~& $E$ (MeV) \\
\hline
0 & $0^+$ & 1 & 0.0 & 0.0 \\
 & $2^+$ & 1 & 7.0 & 3.0 \\
 & $4^+$ & 2 & 23.4, 56.1 & 10.2, 24.3 \\
\hline
1 & $0^-$ & 1 & 30.0 & 13.0 \\
 & $2^+$ & 1 & 44.6 & 19.3 \\
 & $2^-$ & 2 & 37.1, 46.3 & 16.1, 20.1 \\
 & $3^+$ & 1 & 51.6 & 22.4 \\
 & $3^-$ & 1 & 53.3 & 23.1 \\
 & $4^+$ & 1 & 61.0 & 26.5 \\
 & $4^-$ & 3 & 53.5, 62.7, 86.2 & 23.2, 27.2, 37.4 \\
\hline
2 & $0^+$ & 2 & 87.6, 93.5 & 38.0, 40.6 \\
 & $0^-$ & 1 & 90.9 & 39.4 \\
 & $2^+$ & 3 & 94.6, 100.5, 108.1 & 41.0, 43.6, 46.9 \\
 & $2^-$ & 2 & 98.0, 103.0 & 42.5, 44.7 \\
\hline
\end{tabular}
\caption{Energy levels of the quantized $B=8$ Skyrmion. $n$ is the
  number of FR-allowed states with given $I$ and $J^{\pi}$.}
\end{table} 

\begin{figure}[h!]
\begin{center}
\includegraphics[width=14cm]{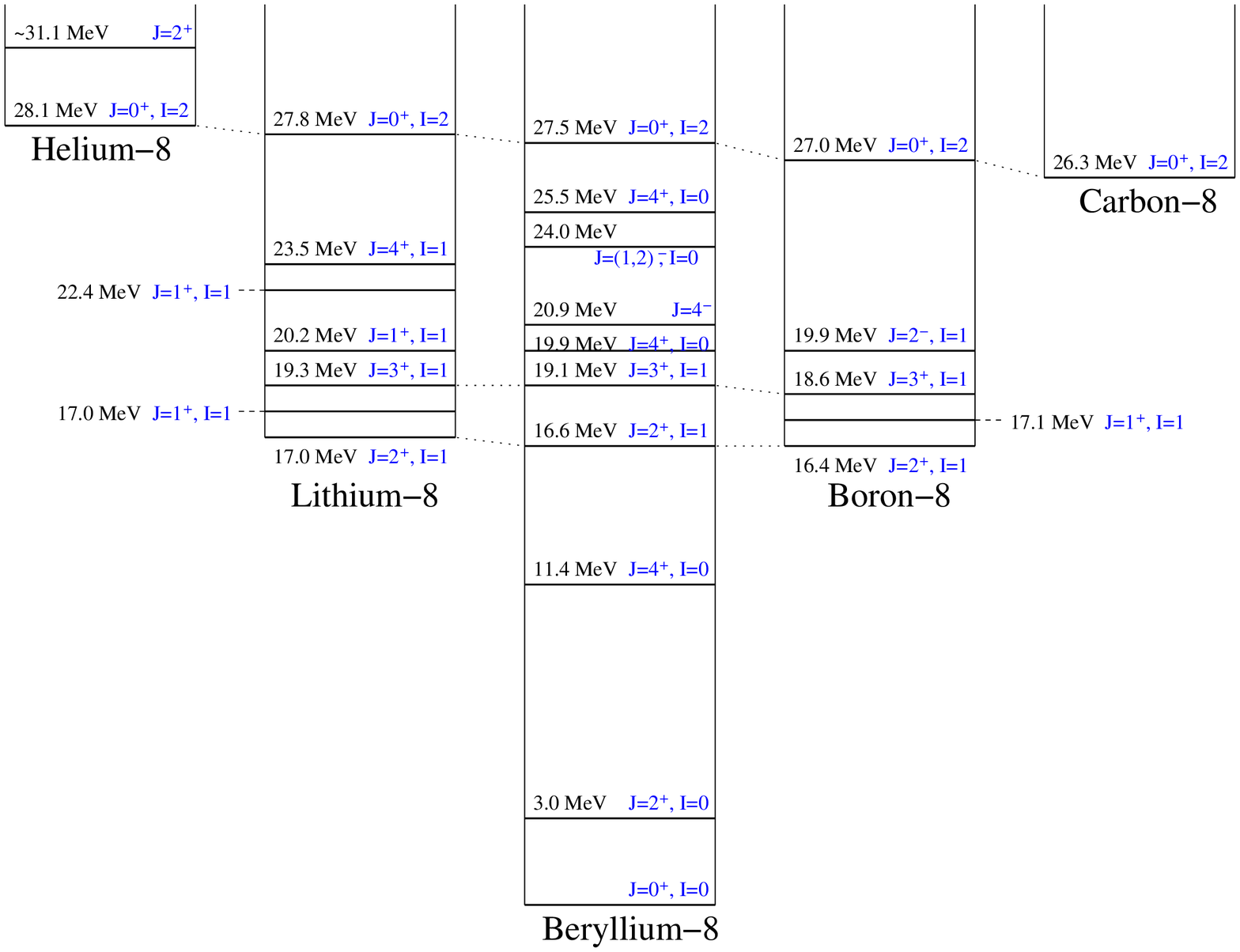}
\caption{Energy level diagram for nuclei of mass number 8. Mass splittings between nuclei are adjusted to eliminate the proton/neutron mass difference and
remove Coulomb effects, as described in Ref. \cite{tilley8910}.}
\end{center}
\end{figure}

Figure 4 is an energy level diagram for nuclei of mass number 8.
The Skyrme model predictions for positive parity states agree well 
with experiment. The ground state of beryllium-8 is correctly determined to be a $0^+$ state.
The rotational band of beryllium-8 is remarkably well reproduced in our model:
we predict the $2^+$ and $4^+$ states at 3.0\,MeV and 10.2\,MeV respectively, which is very close to the experimental values of 3.0\,MeV and 11.4\,MeV respectively.
We predict a second $4^+$ state with isospin 0 at 24.3\,MeV. Experimentally, beryllium-8 has two further $4^+$ states with isospin 0, at 19.9\,MeV and 25.5\,MeV.

The model predicts that the lowest allowed state with isospin 1 has $J^\pi = 0^-$. However, the ground states of lithium-8 and boron-8 are believed to have $J^\pi = 2^+$.
Perhaps the $0^-$ isotriplet may be seen experimentally in the future, for it is known that low-lying spin 0, negative parity states are difficult to observe, 
as experienced in the search for
the bottomonium and charmonium ground state mesons $\eta_b$ and $\eta_c$ \cite{aubert,partridge}. 
Our estimate for the excitation energy of the $2^+$ isotriplet is 19.3\,MeV, which experimentally has an average excitation energy
of 16.7\,MeV. In addition, we 
predict a $3^+$ isotriplet at 22.4\,MeV, to be compared to the experimental value of 19.0\,MeV.
The model forbids spin 1 states with isospin 1. However several $1^+$ states with isospin 1 have been observed in the spectrum of lithium-8. Perhaps these may arise as quantum states
of an alternative $B=8$ Skyrmion, or from the quantization of further degrees of freedom, such as vibrational modes. We predict two $2^-$ states with isospin 1, at 16.1\,MeV and 20.1\,MeV.
A state with these quantum numbers has been seen in the spectrum of boron-8, at 19.9\,MeV.
We predict a $4^+$ state with isospin 1 at 26.5\,MeV, which is seen in the spectrum of  lithium-8 at 23.5\,MeV, 
although its beryllium-8 and boron-8 partners are not yet confirmed. Beryllium-8 has a $4^-$ state at 20.9\,MeV, which we identify with our $4^-$ state with isospin 1 at 23.2\,MeV.

We find that the lowest state with isospin 2 has $J^\pi = 0^+$. 
Experimentally this forms an isospin 2 quintet which includes the ground states of helium-8 and carbon-8. 
We calculate its excitation energy to be 38.0\,MeV, to be compared to the experimental value of 27.3\,MeV. A $2^+$ state of helium-8 has been observed at approximately 31.1\,MeV, which we predict at 41.0\,MeV.

\emph{In summary:} The spectrum of beryllium-8 is very well reproduced in our model, in particular its isospin 0 rotational band. So too are the experimentally observed isospin 1
triplets and the isospin 2 quintet. Predicted $0^-$ states with isospin 1 have not been seen in the spectra of nuclei of mass number 8, however as we mentioned they may be difficult to observe. We have been unable to explain the appearance of $1^+$ states of lithium-8 and boron-8. We are led to consider refinements of our model and its quantization in order to address this.

\section{$B=10$}
In Ref. \cite{battyemas} the minimal-energy solution for $B=10$ was found to have $D_{2h}$
symmetry. This Skyrmion can be thought of within the context 
of the alpha-particle model as a pair of $B=4$ 
cubes with two $B=1$ Skyrmions between, as illustrated in Fig. 6.

\begin{figure}[h!]
\begin{center}
\includegraphics[width=10cm]{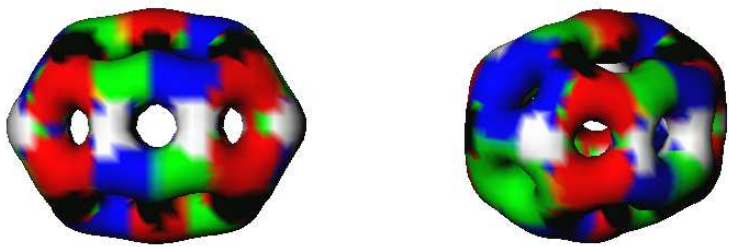}
\caption{A surface of constant baryon density for the $B=10$ Skyrmion (two
viewpoints).}
\end{center}
\end{figure}

Here, we quantize this Skyrmion for the first time.
We use the rational map ansatz to determine its FR constraints.
For an overview of the ansatz, which provides close approximations to the exact Skyrmion
solutions, see Ref. \cite{houghton}. While it does not provide quantitatively exact results, it
precisely describes the symmetry group of many Skyrmions, and therefore
can be used to derive their FR constraints.
In the $B=10$ case a suitable rational map is
\begin{equation} \label{eq:b10map}
R(z)= \frac{\alpha+\beta z^2+\gamma z^4+\delta z^6+\epsilon z^8+z^{10}}{1+\epsilon z^2+\delta z^4+\gamma z^6+\beta z^8+\alpha z^{10}}\,,
\end{equation}
with $\alpha=0.28$, $\beta=-9.37$, $\gamma=14.83$, $\delta=4.98$ and $\epsilon=3.02$. 
The $D_2$ rotation group is generated by two $C_2$ 
symmetries, which correspond to the following symmetries of the rational map:
\begin{equation}
R(-z)=R(z)\,,\,\,\,\,\,R(1/z)=1/R(z)\,.
\end{equation}
The associated FR constraints are determined using 
Krusch's method \cite{krusch} to be
\begin{equation}
e^{i\pi L_3}|\Psi\rangle = |\Psi\rangle\,,\,\,\,\,\,e^{i\pi L_1}e^{i\pi
K_1}|\Psi\rangle = -|\Psi\rangle\,,
\end{equation}
so the signs $\chi_{\rm{FR}}$ form one of the non-trivial 
one-dimensional representations of $D_2$. 

Generally, the parity operator in the Skyrme model is an inversion in space and isospace:
${\cal{P}}: U(\mathbf{x}) \rightarrow U^{\dag}(-\mathbf{x})$. A
rational map is invariant under parity if it satisfies
\begin{equation} 
R(-1/\bar{z}) = -1/\overline{R(z)} \,.
\end{equation}
For a quantized Skyrmion described by such a rational map, all states would have positive parity.
However, as the rational map (\ref{eq:b10map}) satisfies the inversion symmetry
\begin{equation} 
R(-1/\bar{z}) = 1/\overline{R(z)} \,,
\end{equation}
the parity operation in this case is equivalent to a single rotation in isospace, given by 
${\mathcal{P}}=e^{i\pi K_3}$. The parity of a quantum state is its eigenvalue
when acted upon by this ${\mathcal{P}}$. Note that we attach the
parity label to the spin quantum number, to form $J^\pi$, as
is conventionally done in nuclear physics, despite the fact 
that in the Skyrme model the parity operator generally 
reduces to a combination of rotations in space and in isospace.

For the $B=10$ Skyrmion, the symmetries imply that the inertia
tensors $U_{ij}$ and $V_{ij}$ are diagonal, and the only non-zero
component of the mixed inertia tensor $W_{ij}$ is $W_{33}$. 
The quantum Hamiltonian is that of a system of coupled asymmetric tops:
\begin{equation}\label{eq:t10}
H = \frac{1}{2V_{11}}L_1^2
+\frac{1}{2V_{22}}L_2^2
+\frac{U_{33}}{2\Delta_{33}}L_3^2
+\frac{1}{2U_{11}}K_1^2
+\frac{1}{2U_{22}}K_2^2
+\frac{V_{33}}{2\Delta_{33}}K_3^2
+\frac{W_{33}}{\Delta_{33}}L_3K_3\,,
\end{equation}
where $\Delta_{33}=U_{33}V_{33}-W_{33}^2$ as before.
In the absence of symmetry and FR constraints, $(2J+1)\times (2I+1)$ different
non-degenerate levels would correspond to any given pair of $J$ and
$I$. Imposing the FR constraints, however, substantially reduces the number of allowed states.
The calculation of energy levels is similar to the case of a
general asymmetric top, described in Ref. \cite{landau}. However, the final term in (\ref{eq:t10}) mixes states of
the form $|J,L_3\rangle + |J,-L_3\rangle$ and $|J,L_3\rangle - |J,-L_3\rangle$ (and similarly for
isospin basis states). In Table 5 we list the total number of allowed states, $n$, for different
combinations of spin and isospin, together with their energy
eigenvalues. The energy levels are calculated
by diagonalizing the Hamiltonian in matrix form, separately for each combination of 
$J^\pi$ and $I$. The precise forms of the eigenvectors are omitted 
as they do not add anything to our discussion.

\begin{table}[ht]
\centering
\begin{tabular}{|l|l|l|l|l|}
\hline
$I$ & $J^\pi$ & $n$ & $E$ ($\times 10^{-4}$) ~~~& $E$ (MeV) \\
\hline
0 & $1^+$ & 1 & 2.5 & 1.1 \\
 & $2^+$ & 1 & 6.6 & 2.9 \\
 & $3^+$ & 2 & 12.7, 16.1 & 5.5, 7.0 \\
 & $4^+$ & 2 & 21.1, 24.3 & 9.2, 10.6 \\
\hline
1 & $0^+$ & 1 & 23.8 & 10.4 \\
 & $0^-$ & 1 & 24.9 & 10.8 \\
 & $1^-$ & 1 & 27.5 & 11.9 \\
 & $2^+$ & 2 & 30.0, 31.7 & 13.1, 13.8 \\
 & $2^-$ & 3 & 31.1, 31.6, 32.8 & 13.5, 13.8, 14.3 \\
 & $3^+$ & 1 & 37.8 & 16.5 \\
 & $3^-$ & 3 & 37.6, 38.9, 41.1 & 16.4, 16.9, 17.9 \\
 & $4^+$ & 3 & 44.4, 46.1, 51.1 & 19.3, 20.1, 22.2 \\
 & $4^-$ & 5 & 45.5, 46.2, 47.2, 49.3, & 19.8, 20.1, 20.5, 21.5, \\ 
 &       &   & 52.2 & 22.7 \\
\hline
2 & $0^+$ & 1 & 76.0 & 33.1 \\
 & $0^-$ & 1 & 72.7 & 31.6 \\
 & $1^+$ & 2 & 73.9, 78.5 & 32.2, 34.2 \\
 & $1^-$ & 1 & 74.0 & 32.2 \\
 & $2^+$ & 4 & 78.1, 82.2, 82.7, 83.9 & 34.0, 35.8, 36.0, 36.5 \\
 & $2^-$ & 3 & 78.9, 79.2, 80.6 & 34.3, 34.4, 35.0 \\
 & $3^+$ & 5 & 84.2, 87.6, 88.7, 90.1, & 36.6, 38.1, 38.6, 39.2, \\
 &       &   & 92.1 & 40.1 \\
 & $3^-$ & 3 & 85.2, 86.7, 88.6 & 37.1, 37.7, 38.5 \\
\hline
3 & $0^+$ & 2 & 142.9, 147.5 & 62.2, 64.2 \\
 & $0^-$ & 2 & 144.1, 153.2 & 62.7, 66.7 \\
 & $1^+$ & 1 & 150.0 & 65.2 \\
 & $1^-$ & 2 & 146.8, 155.7 & 63.9, 67.7 \\
 & $2^+$ & 5 & 149.1, 150.8, 153.7, 154.2, & 64.9, 65.6, 66.9, 67.1, \\
 &       &   & 155.3 & 67.6 \\
 & $2^-$ & 6 & 150.0, 151.0, 151.7, 159.3, & 65.3, 65.7, 66.0, 69.3, \\
 &       &   & 159.9, 161.1 & 69.6, 70.1 \\
 & $3^+$ & 4 & 156.9, 160.2, 161.6, 163.6 & 68.3, 69.7, 70.3, 71.2 \\
 & $3^-$ & 6 & 157.0, 157.9, 160.4, 165.9, & 68.3, 68.7, 69.8, 72.2, \\
 &       &   & 167.2, 169.3 & 72.8, 73.7 \\
\hline
\end{tabular}
\caption{Energy levels of the quantized $B=10$ Skyrmion.}
\end{table}

The ground state of boron-10 has $J^\pi = 3^+$ and its first
excited state has $J^\pi = 1^+$ at 0.7\,MeV. We incorrectly determine
the $1^+$ state to be the ground state, and $3^+$ states to be excited states.
However, 
this problem arises in other models of boron-10, for example in models involving
nucleon-nucleon potentials in chiral perturbation theory \cite{navratil}.
Boron-10 has further isospin 0 excited states, including $2^+$, $3^+$ and $4^+$
states at
3.6\,MeV, 4.8\,MeV and 6.0\,MeV respectively. We find that our model only allows positive parity states with isospin 0, and our predictions for the excitation energies 
of the aforementioned states are of the correct order of
magnitude. Second $3^+$ and $4^+$ states with isospin 0 are allowed, which we identify with the states of boron-10 at 7.0\,MeV and 10.8\,MeV, respectively.
Curiously, $2^-$, $3^-$ and $4^-$ states of boron-10 with isospin 0 have been found experimentally, at 5.1\,MeV, 6.1\,MeV and 6.6\,MeV, respectively. Consideration of
further degrees of freedom or the quantization of an alternative $B=10$ Skyrmion may be necessary to account for these negative parity states.

We find that the lowest allowed state with isospin 1 has $J^\pi = 0^+$, and is at 4.9\,MeV above the lowest $3^+$ isospin 0 state. Experimentally this is observed as an isospin 1
triplet that includes the beryllium-10 and carbon-10 ground states, and has an average excitation energy of 1.8\,MeV relative to the boron-10 ground state. 
Three $2^+$ spin excitations
of this isotriplet have been observed, at average excitation energies 5.2\,MeV, 7.4\,MeV and 8.9\,MeV. We predict two such excitations at 7.6\,MeV and 8.3\,MeV. 
Note that our model disallows a $J^\pi = 1^+$ state with isospin 1. This agrees with experimental observations. Additionally, $1^-$, $2^-$ and $3^-$ states with isospin 1 have been seen
in the spectrum of beryllium-10, and our predictions for their excitation energies are close to the experimental values. A number of $2^-$ states with isospin 1 are also seen in the spectrum of boron-10.

For isospins 2 and 3, we find that all possible spins and parities are
allowed, and that in both cases the lowest state is a $0^+$ state. The
spins of the lithium-10 and nitrogen-10 states
(both with isospin 2) are not well established. The lowest state of lithium-10 is at 23.3\,MeV above that of boron-10, and its spin
is uncertain: either $1^-$ or $2^-$. We calculate these states to exist at 26.7\,MeV and 28.8\,MeV above the lowest $3^+$ isospin 0 state.
Lithium-10 has an excited $1^+$ state, and the lowest state of nitrogen-10 is believed to be a $1^+$ state. We predict two $1^+$ states with isospin 2, and slightly overpredict
their excitation energies. The lowest helium-10 state is a $0^+$ state at 39.4\,MeV above the boron-10 ground state. This is to be compared to our value of 62.2\,MeV.
In our model, the isobar splittings increase in proportion to $I(I+1)$. However,
as can be seen from the nuclear energy level diagrams, this approximately quadratic behaviour is not
precisely reflected in the data.

\emph{In summary:} The quantum numbers and excitation energies of the states of nuclei of mass number 10 are reasonably well described by our model. However, the appearance of a
$1^+$ state as the isospin 0 ground state disagrees with that of boron-10, but as we mentioned this is a well-known artifact of nuclear models. 
The negative parity states with isospin 0 in the nuclear spectra may arise as quantum states of an alternative $B=10$ Skyrmion, invariant under an alternative symmetry group which
may permit such states. We find significantly more allowed states with isospins 2 and 3 than have been found experimentally. However we are not too concerned about those of higher spin, as the $D_{2h}$ symmetry may become deformed as the Skyrmion spins.

\begin{figure}[h!]
\begin{center}
\includegraphics[width=15.5cm]{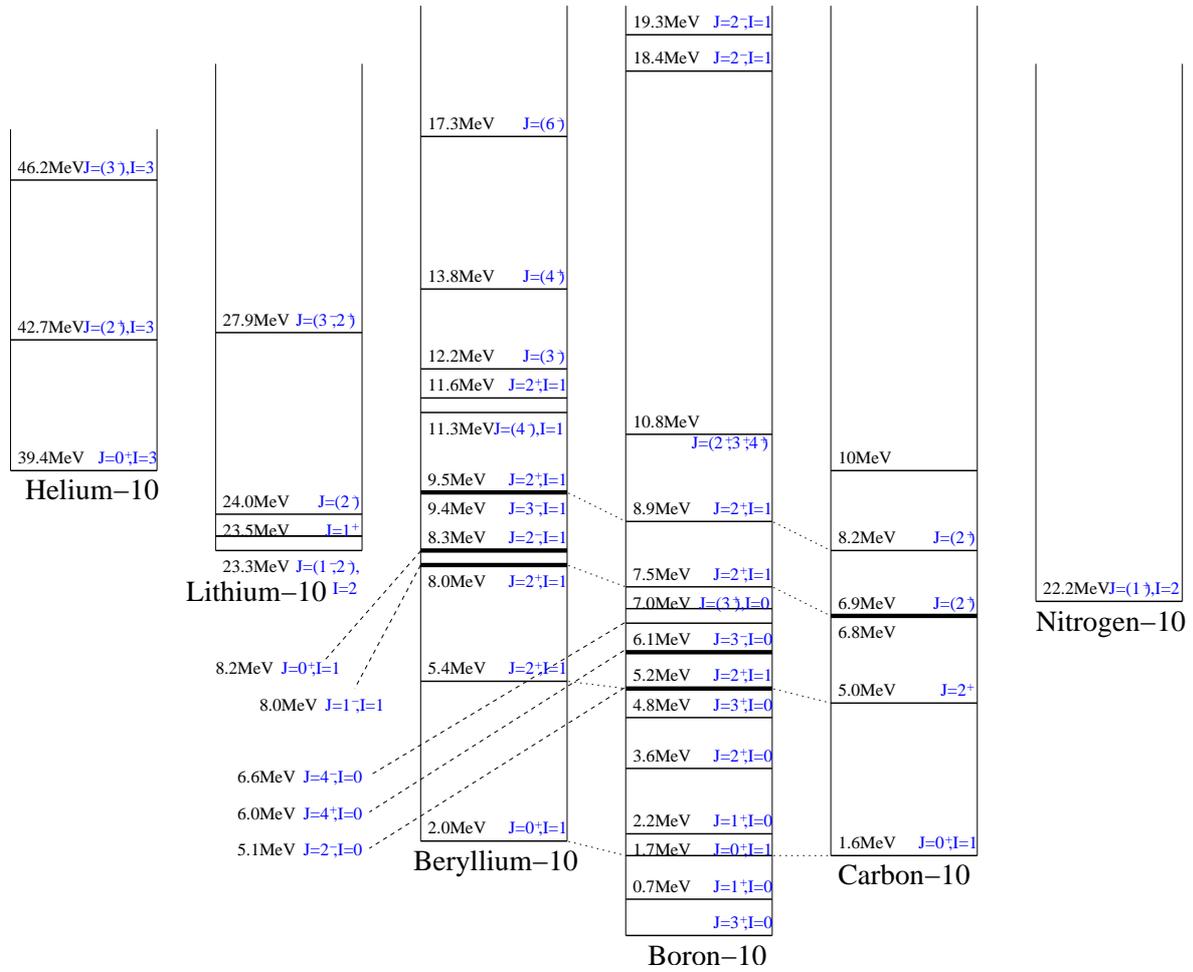}
\caption{Energy level diagram for nuclei of mass number 10. Individual isobars 
are shifted vertically for clarity, and mass splittings between nuclei are adjusted to 
eliminate the proton/neutron mass difference and remove Coulomb
effects, as described in Ref. \cite{tilley8910}.}
\end{center}
\end{figure}

\section{$B=12$}
In the alpha-particle model, the classical minimum of the potential energy for three alpha-particles 
occurs when they are arranged in an equilateral triangle. The minimal-energy solution 
of the Skyrme model in the $B=12$ sector has $C_{3v}$ symmetry,
but there is a solution of very slightly higher energy with a larger $D_{3h}$ symmetry, which is a saddle
point and not a local energy minimum. Both have an equilateral
triangle shape. Here, we quantize the $D_{3h}$-symmetric solution (see Fig. 8), which we believe
provides a physically more realistic picture of the nucleus. A more
refined analysis might include an anharmonic  vibrational mode centred on the
$D_{3h}$-symmetric solution and oscillating through
two $C_{3v}$-symmetric minima. This solution can be approximated using the
double rational map ansatz \cite{mp}. We use this to determine its FR constraints.
The ansatz involves a $D_{3h}$-symmetric outer map of degree 11, $R^{\rm{out}}$, and a
spherically symmetric degree 1 inner map, $R^{\rm{in}}$, together with an overall radial profile function.
The maps are \cite{bms}:
\begin{eqnarray}
R^{\rm{out}}(z) &=& \frac{z^9+\alpha z^6+\beta z^3+\gamma}{z^2(\gamma z^9+\beta z^6+\alpha z^3+1)}\,,\\
R^{\rm{in}}(z) &=& -z\,,
\end{eqnarray}
where $\alpha=-2.47$, $\beta=-0.84$ and $\gamma=-0.13$. The orientation of the inner map is chosen
so that the $D_{3h}$ symmetry is realised in a way compatible with the outer map. Both maps
satisfy
\begin{equation}
R(e^{i\frac{2\pi}{3}}z)=e^{i\frac{2\pi}{3}}R(z)\,,\,\,\,\,\,R(1/z)=1/R(z)\,.
\end{equation}
As the baryon number is a multiple of four, the FR signs form the trivial representation of
$D_3$, and so the FR constraints are \cite{krusch}
\begin{equation}
e^{i\frac{2\pi}{3}L_3}e^{i\frac{2\pi}{3}K_3}|\Psi\rangle =|\Psi\rangle \,,\,\,\,\,\,
e^{i\pi L_1}e^{i\pi K_1} |\Psi\rangle =|\Psi\rangle\,.
\end{equation}
Both maps satisfy the reflection symmetry
\begin{equation}
R(1/\bar{z})=1/\overline{R(z)}\,,
\end{equation}
and so the parity operator is equivalent to ${\mathcal{P}}=e^{i\pi L_3}e^{i\pi K_3}$\,. The $D_{3h}$
symmetry implies that the inertia tensors are diagonal, with $U_{11}=U_{22}$, $V_{11}=V_{22}$
and $W_{11}=W_{22}$, so the quantum Hamiltonian is that of a system of coupled symmetric tops:
\begin{equation*}
H=\left(\frac{U_{11}-W_{11}}{2\Delta_{11}}\right)\mathbf{J}^2
+\left(\frac{V_{11}-W_{11}}{2\Delta_{11}}\right)\mathbf{I}^2
+\left(\frac{W_{11}}{2\Delta_{11}}\right)\mathbf{M}^2
\end{equation*}
\begin{equation}
+\left(\frac{U_{33}}{2\Delta_{33}}-\frac{U_{11}}{2\Delta_{11}}\right)L_3^2
+\left(\frac{V_{33}}{2\Delta_{33}}-\frac{V_{11}}{2\Delta_{11}}\right)K_3^2
+\left(\frac{W_{33}}{\Delta_{33}}-\frac{W_{11}}{\Delta_{11}}\right)L_3K_3\,,
\end{equation}
where $\mathbf{M}=\mathbf{L}+\mathbf{K}$, $\Delta_{33}$ is
as before, and $\Delta_{11} = U_{11}V_{11}-W_{11}^2$.

\begin{figure}[h!]
\begin{center}
\includegraphics[width=10cm]{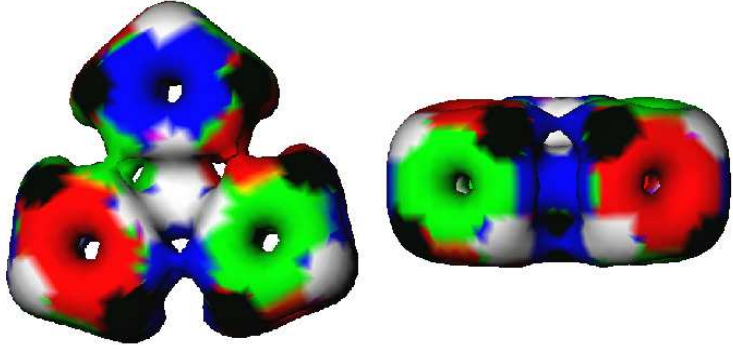}
\caption{A surface of constant baryon density for the $D_{3h}$-symmetric $B=12$
Skyrmion (two viewpoints).}
\end{center}
\end{figure}

The states that are allowed by the FR constraints are listed in Table 6.
Each of the allowed isospin 0 states is also an eigenstate of the Hamiltonian, and
their quantum energies $E_{J^\pi,I,|L_3|,|K_3|}$ are:
\begin{eqnarray}
E_{0^+,0,0,0} &=& 0\,, \nonumber\\
E_{2^+,0,0,0} &=& 3U_{11}/\Delta_{11}\,, \nonumber\\
E_{3^-,0,3,0} &=& 3U_{11}/2\Delta_{11}+9U_{33}/2\Delta_{33}\,, \nonumber\\
E_{4^-,0,3,0} &=& 11U_{11}/2\Delta_{11}+9U_{33}/2\Delta_{33}\,, \nonumber\\
E_{4^+,0,0,0} &=& 10U_{11}/\Delta_{11}\,, \nonumber\\
E_{5^-,0,3,0} &=& 21U_{11}/2\Delta_{11}+9U_{33}/2\Delta_{33}\,, \nonumber\\
E_{6^-,0,3,0} &=& 33U_{11}/2\Delta_{11}+9U_{33}/2\Delta_{33}\,, \nonumber\\
E_{6^+,0,0,0} &=& 21U_{11}/\Delta_{11}\,, \nonumber\\
E_{6^+,0,6,0} &=& 3U_{11}/\Delta_{11}+18U_{33}/\Delta_{33}\,.
\end{eqnarray}
These states also result from the rigid-body quantization of an equilateral 
triangle with alpha-particles at its vertices, and are not a
prediction characteristic of the Skyrme model itself. The states fall into
rotational bands labelled by $|L_3|=0,\,3,\,6,\,...$. As in
Refs. \cite{alvarez,alvarez1}, we suggest that the second experimental $2^-$ state of carbon-12 at 13.4\,MeV
has been misidentified, and is really a $4^-$ state. 
Again as in Ref. \cite{alvarez}, we predict a relatively low-energy $6^+$ state of carbon-12,
with $|L_3|=6$. Such a state has not yet been seen experimentally.

The isospin 1 states in Table 6 are not all individually
eigenstates of the Hamiltonian, as they are not eigenstates of 
$\mathbf{M}^2$. It is convenient to introduce basis states
$|J,I;M,M_3\rangle$, and express the states we have in terms of these.
For example, for the two orthonormal FR-allowed $1^+$ states:
\begin{eqnarray}
\left|\Psi_{1^+,1,0,0} \right\rangle &\equiv & 
|1,0\rangle \otimes |1,0\rangle =
\sqrt{\frac{2}{3}}|1,1;2,0\rangle-\sqrt{\frac{1}{3}}|1,1;0,0\rangle\,,\\
\left|\Psi_{1^+,1,1,1} \right\rangle &\equiv &
\frac{1}{\sqrt{2}}\left(|1,1\rangle \otimes |1,-1\rangle +
|1,-1\rangle \otimes |1,1\rangle \right) =
\sqrt{\frac{1}{3}}|1,1;2,0\rangle+\sqrt{\frac{2}{3}}|1,1;0,0\rangle\,.
\end{eqnarray}
Neither is an eigenstate of the Hamiltonian, as neither is an eigenstate of $\mathbf{M}^2$, 
but we can find two linear combinations of $\left|\Psi_{1^+,1,0,0}
\right\rangle$ and $\left|\Psi_{1^+,1,1,1} \right\rangle$ that are
eigenstates by diagonalizing a matrix Hamiltonian. 
Their energies are the eigenvalues of the $2\times 2$ matrix
\begin{equation} \label{eq:matrix}
\left(
\begin{array}{cc}
\left\langle\Psi_{1^+,1,0,0} \right|H\left|\Psi_{1^+,1,0,0} \right\rangle & \left\langle\Psi_{1^+,1,0,0} \right|H\left|\Psi_{1^+,1,1,1} \right\rangle \\
\left\langle\Psi_{1^+,1,1,1} \right|H\left|\Psi_{1^+,1,0,0} \right\rangle & \left\langle\Psi_{1^+,1,1,1} \right|H\left|\Psi_{1^+,1,1,1} \right\rangle
\end{array}
\right)
\equiv
\left(
\begin{array}{cc}
a & c \\
c & b
\end{array}
\right)\,,
\end{equation}
where
\begin{eqnarray}
a &=& U_{11}/\Delta_{11}+V_{11}/\Delta_{11}\,,\\
b &=&
U_{11}/2\Delta_{11}+V_{11}/2\Delta_{11}+U_{33}/2\Delta_{33}+V_{33}/2\Delta_{33}-W_{33}/\Delta_{33}\,,\\
c &=& \sqrt{2}W_{11}/\Delta_{11}\,.
\end{eqnarray}
The eigenvalues are $\frac{1}{2}(a+b \pm
\sqrt{a^2+b^2-2ab+4c^2})$. There are two possible interpretations.
Either the energies are close together and remain
below the isospin 1 states of higher spin, in which case we predict
two close $1^+$ states (which have experimentally not been resolved), 
or their energies are well separated, in which case we predict the
observed $1^+$ state and a higher excited $1^+$ state that
has not yet been seen. From our numerical values for $a,b$ and $c$ we 
calculate the energies to be 0.00194 and 0.00207, which are in fact
close together. So we predict that the single observed $1^+$ isotriplet of states is really an unresolved 
doublet of isotriplets.

This matrix diagonalization method is also used to calculate the 
energy eigenvalues for the other values of $J^\pi$ and $I$  
for which there is more than one allowed state. It is found that 
in all cases, the off-diagonal elements of the matrices analogous to
(\ref{eq:matrix}) are small, of the order $10^{-2}$ times the 
diagonal elements. We may therefore consistently assign values of $|L_3|$ and $|K_3|$ to
each of our calculated energy eigenvalues, as the `mixing' of states
with the same values of $J^\pi$ and $I$, but different $|L_3|$ and
$|K_3|$ values, is minimal. 
The quantum energies of the spin 2 states with isospin 1 are:
\begin{eqnarray}
E_{2^-,1,2,1} &=& U_{11}/\Delta_{11}+2U_{33}/\Delta_{33}+V_{11}/2\Delta_{11}+V_{33}/2\Delta_{33}+2W_{33}/\Delta_{33}\,,\\
E_{2^+,1,1,1} &=& 5U_{11}/2\Delta_{11}+U_{33}/2\Delta_{33}+V_{11}/2\Delta_{11}+V_{33}/2\Delta_{33}-W_{33}/\Delta_{33}\,.
\end{eqnarray}
We calculate that $E_{2^-,1,2,1}$ and $E_{2^+,1,1,1}$ are 0.00218 and 0.00228 respectively, and so 
the $2^+$ isotriplet lies above the $2^-$ isotriplet. Experimentally, however, the $2^+$ isotriplet is observed below the $2^-$ isotriplet.

The quantum energies of the isospin 2 states with spins 0 and 1 are:
\begin{eqnarray}
E_{0^+,2,0,0} &=& 3V_{11}/\Delta_{11}\,,\\
E_{1^-,2,1,2} &=& U_{11}/2\Delta_{11}+U_{33}/2\Delta_{33}+V_{11}/\Delta_{11}+2V_{33}/\Delta_{33}+2W_{33}/\Delta_{33}\,,\\
E_{1^+,2,1,1} &=& U_{11}/2\Delta_{11}+U_{33}/2\Delta_{33}+5V_{11}/2\Delta_{11}+V_{33}/2\Delta_{33}-W_{33}/\Delta_{33}\,.
\end{eqnarray}
We calculate these values to be 0.00569, 0.00541 and 0.00574
respectively, and so the $1^-$ state lies below the $0^+$ state,
and the $1^+$ state lies above the $0^+$ state. 
Between the $0^+$ and $1^+$ states there is a further $2^+$ state.
Experimentally 
the $0^+$ isospin 2 quintet includes the ground states of beryllium-12 and oxygen-12; low-energy excited $1^-$ and $2^+$ states of beryllium-12 are also observed.
The energy levels of the quantized $B=12$ Skyrmion are listed in Table
7. The experimental spectrum is in Fig. 9.

\begin{table}[ht]
\centering
\begin{tabular}{|l|l|l|}
\hline
$I$ & $J^\pi$ & Wave function $\left|\Psi_{J^\pi,I,|L_3|,|K_3|} \right\rangle$ ~~~\\
\hline
0 & $0^+$ & $\left|\Psi_{0^+,0,0,0} \right\rangle = |0,0\rangle \otimes |0,0\rangle$ \\
  & $2^+$ & $\left|\Psi_{2^+,0,0,0} \right\rangle = |2,0\rangle \otimes |0,0\rangle$ \\
  & $3^-$ & $\left|\Psi_{3^-,0,3,0} \right\rangle = \frac{1}{\sqrt{2}}(|3,3\rangle-|3,-3\rangle) \otimes |0,0\rangle$ \\
  & $4^-$ & $\left|\Psi_{4^-,0,3,0} \right\rangle = \frac{1}{\sqrt{2}}(|4,3\rangle+|4,-3\rangle) \otimes |0,0\rangle$ \\
  & $4^+$ & $\left|\Psi_{4^+,0,0,0} \right\rangle = |4,0\rangle \otimes |0,0\rangle$ \\
  & $5^-$ & $\left|\Psi_{5^-,0,3,0} \right\rangle = \frac{1}{\sqrt{2}}(|5,3\rangle-|5,-3\rangle) \otimes |0,0\rangle$ \\
  & $6^-$ & $\left|\Psi_{6^-,0,3,0} \right\rangle = \frac{1}{\sqrt{2}}(|6,3\rangle+|6,-3\rangle) \otimes |0,0\rangle$ \\
  & $6^+$ & $\left|\Psi_{6^+,0,0,0} \right\rangle = |6,0\rangle \otimes |0,0\rangle$ \\
  &       & $\left|\Psi_{6^+,0,6,0} \right\rangle = \frac{1}{\sqrt{2}}(|6,6\rangle+|6,-6\rangle) \otimes |0,0\rangle$ \\
\hline
1 & $1^+$ & $\left|\Psi_{1^+,1,1,1} \right\rangle = \frac{1}{\sqrt{2}}\left(|1,1\rangle \otimes |1,-1\rangle + |1,-1\rangle \otimes |1,1\rangle \right)$ \\
  &       & $\left|\Psi_{1^+,1,0,0} \right\rangle = |1,0\rangle \otimes |1,0\rangle$ \\
  & $2^-$ & $\left|\Psi_{2^-,1,2,1} \right\rangle = \frac{1}{\sqrt{2}}\left(|2,2\rangle \otimes |1,1\rangle - |2,-2\rangle \otimes |1,-1\rangle \right)$ \\
  & $2^+$ & $\left|\Psi_{2^+,1,1,1} \right\rangle = \frac{1}{\sqrt{2}}\left(|2,1\rangle \otimes |1,-1\rangle - |2,-1\rangle \otimes |1,1\rangle \right)$ \\
  & $3^-$ & $\left|\Psi_{3^-,1,3,0} \right\rangle = \frac{1}{\sqrt{2}}(|3,3\rangle+|3,-3\rangle) \otimes |1,0\rangle$ \\
  &       & $\left|\Psi_{3^-,1,2,1} \right\rangle = \frac{1}{\sqrt{2}}\left(|3,2\rangle \otimes |1,1\rangle + |3,-2\rangle \otimes |1,-1\rangle \right)$ \\
  & $3^+$ & $\left|\Psi_{3^+,1,1,1} \right\rangle = \frac{1}{\sqrt{2}}\left(|3,1\rangle \otimes |1,-1\rangle + |3,-1\rangle \otimes |1,1\rangle \right)$ \\
  &       & $\left|\Psi_{3^+,1,0,0} \right\rangle = |3,0\rangle \otimes |1,0\rangle$ \\
  & $4^-$ & $\left|\Psi_{4^-,1,4,1} \right\rangle = \frac{1}{\sqrt{2}}\left(|4,4\rangle \otimes |1,-1\rangle - |4,-4\rangle \otimes |1,1\rangle \right)$ \\
  &       & $\left|\Psi_{4^-,1,3,0} \right\rangle = \frac{1}{\sqrt{2}}(|4,3\rangle-|4,-3\rangle) \otimes |1,0\rangle$ \\
  &       & $\left|\Psi_{4^-,1,2,1} \right\rangle = \frac{1}{\sqrt{2}}\left(|4,2\rangle \otimes |1,1\rangle - |4,-2\rangle \otimes |1,-1\rangle \right)$ \\
  & $4^+$ & $\left|\Psi_{4^+,1,1,1} \right\rangle = \frac{1}{\sqrt{2}}\left(|4,1\rangle \otimes |1,-1\rangle - |4,-1\rangle \otimes |1,1\rangle \right)$ \\
\hline
2 & $0^+$ & $\left|\Psi_{0^+,2,0,0} \right\rangle = |0,0\rangle \otimes |2,0\rangle$ \\
  & $1^-$ & $\left|\Psi_{1^-,2,1,2} \right\rangle = \frac{1}{\sqrt{2}}\left(|1,1\rangle \otimes |2,2\rangle - |1,-1\rangle \otimes |2,-2\rangle \right)$ \\
  & $1^+$ & $\left|\Psi_{1^+,2,1,1} \right\rangle = \frac{1}{\sqrt{2}}\left(|1,1\rangle \otimes |2,-1\rangle - |1,-1\rangle \otimes |2,1\rangle \right)$ \\
  & $2^-$ & $\left|\Psi_{2^-,2,1,2} \right\rangle = \frac{1}{\sqrt{2}}\left(|2,1\rangle \otimes |2,2\rangle + |2,-1\rangle \otimes |2,-2\rangle \right)$ \\
  &       & $\left|\Psi_{2^-,2,2,1} \right\rangle = \frac{1}{\sqrt{2}}\left(|2,2\rangle \otimes |2,1\rangle + |2,-2\rangle \otimes |2,-1\rangle \right)$ \\
  & $2^+$ & $\left|\Psi_{2^+,2,2,2} \right\rangle = \frac{1}{\sqrt{2}}\left(|2,2\rangle \otimes |2,-2\rangle + |2,-2\rangle \otimes |2,2\rangle \right)$ \\
  &       & $\left|\Psi_{2^+,2,1,1} \right\rangle = \frac{1}{\sqrt{2}}\left(|2,1\rangle \otimes |2,-1\rangle + |2,-1\rangle \otimes |2,1\rangle \right)$ \\
  &       & $\left|\Psi_{2^+,2,0,0} \right\rangle = |2,0\rangle \otimes |2,0\rangle$ \\
\hline
\end{tabular}
\caption{FR-allowed quantum states of the $B=12$ Skyrmion (normalised).}
\end{table}

\begin{table}[ht]
\centering
\begin{tabular}{|l|l|l|l|l|l|l|l|}
\hline
$I$ & $J^\pi$ & $|L_3|$ & $|K_3|$ & $E^{\rm{exact}}$ ~~~& $E^{\rm{3-cube}}$ ~~~& $E^{\rm{exact}}$ ~~~& $E^{\rm{3-cube}}$ ~~~\\
    ~~~&         ~~~&         ~~~&         ~~~& ($\times 10^{-4}$) ~~~& ($\times 10^{-4}$) ~~~& (MeV) ~~~& (MeV) ~~~ \\
\hline
0 & $0^+$ & 0 & 0 & 0.0 & 0.0 & 0.0 & 0.0 \\
  & $2^+$ & 0 & 0 & 5.1 & 5.2 & 3.2 & 3.2 \\
  & $3^-$ & 3 & 0 & 7.6 & 7.6 & 4.7 & 4.7 \\
  & $4^-$ & 3 & 0 & 14.4 & 14.5 & 8.9 & 9.0 \\
  & $4^+$ & 0 & 0 & 17.0 & 17.4 & 10.6 & 10.8 \\
  & $5^-$ & 3 & 0 & 22.9 & 23.2 & 14.2 & 14.4 \\
  & $6^-$ & 3 & 0 & 33.1 & 33.7 & 20.6 & 20.9 \\
  & $6^+$ & 6 & 0 & 25.3 & 25.2 & 15.7 & 15.6 \\
  &       & 0 & 0 & 35.7 & 36.5 & 22.2 & 22.7 \\
 
\hline
1 & $1^+$ & 1 & 1 & 19.4 & 18.4 & 12.1 & 11.5 \\
  &       & 0 & 0 & 20.7 & 20.2 & 12.9 & 12.6 \\
  & $2^-$ & 2 & 1 & 21.8 & 21.0 & 13.5 & 13.0 \\
  & $2^+$ & 1 & 1 & 22.8 & 21.9 & 14.2 & 13.6 \\
  & $3^-$ & 3 & 0 & 26.4 & 26.1 & 16.4 & 16.2 \\
  &       & 2 & 1 & 27.0 & 26.2 & 16.8 & 16.3 \\
  & $3^+$ & 1 & 1 & 27.8 & 27.1 & 17.2 & 16.8 \\
  &       & 0 & 0 & 29.3 & 28.9 & 18.2 & 18.0 \\
  & $4^-$ & 4 & 1 & 30.6 & 29.4 & 19.0 & 18.2 \\
  &       & 3 & 0 & 33.1 & 33.1 & 20.6 & 20.5 \\
  &       & 2 & 1 & 33.9 & 33.1 & 21.1 & 20.6 \\
  & $4^+$ & 1 & 1 & 34.7 & 34.1 & 21.6 & 21.2 \\
\hline
2 & $0^+$ & 0 & 0 & 56.9 & 55.5 & 35.4 & 34.5 \\
  & $1^-$ & 1 & 2 & 54.1 & 50.9 & 33.6 & 31.6 \\
  & $1^+$ & 1 & 1 & 57.4 & 55.4 & 35.7 & 34.4 \\
  & $2^-$ & 1 & 2 & 57.5 & 54.4 & 35.7 & 33.8 \\
  &       & 2 & 1 & 59.7 & 58.0 & 37.1 & 36.0 \\
  & $2^+$ & 2 & 2 & 57.1 & 53.5 & 35.5 & 33.2 \\
  &       & 1 & 1 & 60.6 & 58.9 & 37.7 & 36.6 \\
  &       & 0 & 0 & 62.3 & 60.7 & 38.7 & 37.7 \\
\hline
\end{tabular}
\caption{Energy levels of the $B=12$ Skyrmion, using both the exact solution and the three cube interpretation. To each of the 
quantum states of the exact solution there correspond dominant values of $|L_3|$ and $|K_3|$.}
\end{table}

\subsection{Three cube interpretation} 
In Ref. \cite{mmw} we estimated the moments of inertia of the $B=8$ Skyrmion by treating it as a `double cube' 
of two cubic $B=4$ Skyrmions, separated a certain distance along a common $C_4$ axis, with the cubes
rotated around this axis by $\pi /2$ relative to each other. This enabled us to estimate the energies of the Skyrmion's
quantum states.
These estimates agree well with our results using the
exact $B=8$ solution. In this section we apply a similar procedure to the $B=12$ Skyrmion.

We work with three $B=4$ cubes arranged in an equilateral triangle, meeting at a
common edge. Each cube is related to its neighbour by a spatial rotation by $2\pi/3$
followed by an isorotation by $2\pi/3$. The isorotation cyclically permutes the values of
the pion fields on the faces of the cubes, so that these values match on touching faces.
We denote by $d$ the distance between the centre of each cube and the centre of the
triangle. In a similar manner to the case of the $B=8$
double cube, we use the parallel axis theorem to make estimates for the moments of inertia of the $D_{3h}$-symmetric
$B=12$ Skyrmion in terms of the moments of inertia of the $B=4$
Skyrmion and the separation parameter $d$. We use the quadratic interpolation method to obtain values for the $B=4$ inertia tensors
with $m=0.685$, the optimised value of $m$ for the $B=12$ sector of the model (see Table 1), to obtain the estimates
\begin{eqnarray}
U_{11}^{(B=12)}&=&U_{22}^{(B=12)}=3U_{11}^{(B=4)} =540\,,\\
U_{33}^{(B=12)}&=&3U_{33}^{(B=4)} =645\,,\\
V_{11}^{(B=12)}&=&V_{22}^{(B=12)}=3V_{11}^{(B=4)}+\frac{3}{2}{\cal{M}}_4d^2\,,\\
V_{33}^{(B=12)}&=&3V_{11}^{(B=4)}+3{\cal{M}}_4d^2\,.
\end{eqnarray}
The interpolated value of ${\cal{M}}_4$ for $m=0.685$ is 589.3.
The value of $d$ is chosen using a least squares method so that our approximation to $V_{ij}$ is 
closest to that of the exact $B=12$ solution. This yields a value for $d$ of
1.92, and hence the estimates
\begin{eqnarray}
V_{11}^{(B=12)}&=& 5756\,,\\
V_{33}^{(B=12)}&=& 9014\,.
\end{eqnarray}
In this simplified picture, $W_{ij}$ vanishes, as it vanishes for the $B=4$ cube. Comparing these numbers to those
obtained from the exact solution for $B=12$ (given in Appendix A.5), we see that the three cube approach has
provided good estimates of the inertia tensors $U_{ij}$ and $V_{ij}$, 
and the inequalities that are satisfied by their elements 
are right. For the exact solution, $W_{ij}$ is non-zero but small.
Also, for the exact solution the ratio of $U_{11}$ to $U_{33}$ is closer to 1 than for the $B=4$
cube. This is because the triangular arrangement of the three cubes is closer to the Skyrme crystal, for
which $U_{11}=U_{33}$. The accuracy of this approximate inertia
tensor shows that the Skyrme model is consistent with 
the intrinsic shape of carbon-12 being an equilateral triangle 
of three alpha-particles.

\pagebreak
The assumption that $W_{ij}$ vanishes 
simplifies the quantum Hamiltonian to the sum of a symmetric top in space and a symmetric top in isospace:
\begin{equation}
H=\frac{1}{2V_{11}}{\mathbf{J}}^2
+\frac{1}{2U_{11}}{\mathbf{I}}^2
+\left(\frac{1}{2V_{33}}-\frac{1}{2V_{11}}\right)L_3^2
+\left(\frac{1}{2U_{33}}-\frac{1}{2U_{11}}\right)K_3^2\,.
\end{equation}
$|L_3|$ and $|K_3|$ become good quantum numbers, and the expressions
for a selection of the quantum energies simplify to:
\begin{eqnarray}
E_{0^+,0,0,0} &=& 0\,, \nonumber\\
E_{2^+,0,0,0} &=& 3/V_{11}\,, \nonumber\\
E_{3^-,0,3,0} &=& 3/2V_{11}+9/2V_{33}\,, \nonumber\\
E_{4^-,0,3,0} &=& 11/2V_{11}+9/2V_{33}\,, \nonumber\\
E_{4^+,0,0,0} &=& 10/V_{11}\,, \nonumber\\
E_{5^-,0,3,0} &=& 21/2V_{11}+9/2V_{33}\,, \nonumber\\
E_{6^-,0,3,0} &=& 33/2V_{11}+9/2V_{33}\,, \nonumber\\
E_{6^+,0,0,0} &=& 21/V_{11}\,, \nonumber\\
E_{6^+,0,6,0} &=& 3/V_{11}+18/V_{33}\,, \nonumber\\
E_{1^+,1,1,1} &=& 1/V_{11}+1/U_{11}\,, \nonumber\\
E_{1^+,1,0,0} &=& 1/2V_{11}+1/2V_{33}+1/2U_{11}+1/2U_{33}\,, \nonumber\\
E_{2^-,1,2,1} &=& 1/V_{11}+2/V_{33}+1/2U_{11}+1/2U_{33}\,, \nonumber\\
E_{2^+,1,1,1} &=& 5/2V_{11}+1/2V_{33}+1/2U_{11}+1/2U_{33}\,, \nonumber\\
E_{0^+,2,0,0} &=& 3/U_{11}\,, \nonumber\\
E_{1^-,2,1,2} &=& 1/2V_{11}+1/2V_{33}+1/U_{11}+2/U_{33}\,, \nonumber\\
E_{1^+,2,1,1} &=& 1/2V_{11}+1/2V_{33}+5/2U_{11}+1/2U_{33}\,, \nonumber\\
E_{2^+,2,2,2} &=& 1/V_{11}+2/V_{33}+1/U_{11}+2/U_{33}\,.
\end{eqnarray}
Their numerical values (using $U_{11}=540$, $U_{33}=645$,
$V_{11}=5756$ and $V_{33}=9014$) are listed in Table 7 
alongside the corresponding values using the exact solution.
We also present the values in physical units, obtained using the $B=12$ parameter set given in Table 1.

\subsection{Comparison with experimental data}
The ground state of carbon-12 is a $0^+$ state with isospin 0. It has excited $2^+$, $3^-$ and $4^+$ 
states with excitation energies 4.4\,MeV, 9.6\,MeV and 14.1\,MeV
respectively. In addition, there may be a $4^-$ state
at 13.4\,MeV (reassigned from the $2^-$ state at this energy).
As can be seen from Table 7, we predict precisely these states, 
and in the same order. Our predictions for their excitation energies are 3.2\,MeV, 4.7\,MeV, 10.6\,MeV and 8.9\,MeV respectively.

Carbon-12 has an excited $0^+$ state 
at 7.7\,MeV, the famous Hoyle state. Unfortunately our method 
of rigid-body quantization prohibits two independent spin 0, 
isospin 0 states. An extension of the model might allow 
this. Perhaps the lowest-lying quantum state of an alternative $B=12$ solution, 
such as the $C_{3v}$-symmetric solution, or a solution with 
three $B=4$ Skyrmions in a linear chain, both discussed in 
Ref. \cite{bms}, could be interpreted as this excited $0^+$ state. 
Additional excited states of carbon-12 are seen experimentally 
but are not yet predicted in our model: for example, 
a $1^+$ state at 12.7\,MeV. 

We predict two non-degenerate $1^+$ states with isospin 1. 
An isotriplet with $J^\pi=1^+$ is
observed, which includes the ground states of boron-12 and nitrogen-12, with average excitation energy
15.1\,MeV, to be compared to our value of 12.1\,MeV. As mentioned previously, this may be an unresolved doublet of isotriplets. However, a number of higher $1^+$ states with isospin 1 are seen in the nuclear spectra.
We also predict a $2^+$ and a $2^-$ isotriplet. Both of these are seen experimentally, but in the opposite
order: we predict the $2^-$ isotriplet to lie below the $2^+$ isotriplet. Experimentally, the $2^+$
isotriplet has an average excitation energy of 16.1\,MeV, compared with 16.5\,MeV for the $2^-$ isotriplet.
Higher excited $1^-$ and $0^+$ isotriplets are seen experimentally, but we do not predict
them in our model. 
We find two allowed $3^+$ states with isospin 1, at 16.8\,MeV and 18.0\,MeV. One such state has been seen in the spectrum of
boron-12, at 20.8\,MeV.
An incomplete $3^-$ isotriplet is observed at 18.5\,MeV, which we predict at 16.4\,MeV. We predict a second $3^-$ state with isospin 1 at 16.8\,MeV. States with these quantum numbers are seen in the spectra of boron-12, carbon-12 and nitrogen-12, at 20.9\,MeV, 20.6\,MeV and 20.4\,MeV, respectively.
A further excited $4^-$
state of boron-12 with isospin 1 exists at 19.7\,MeV, to be compared with our value of 19.0\,MeV.

An (incomplete) $J^{\pi}=0^+$, isospin 2 quintet, which includes the ground states of beryllium-12 and
oxygen-12, is observed experimentally with an average excitation energy of 27.7\,MeV. 
We predict such an isoquintet at 35.4\,MeV. We also predict the existence of a $1^-$ isoquintet with an excitation energy 
less than that of the $0^+$ isoquintet. 
Such a $1^-$ state is observed in beryllium-12, at 2.7\,MeV above the $0^+$ ground state.
An (incomplete) $2^+$ isoquintet is experimentally observed with an average excitation energy roughly 2\,MeV
above the $0^+$ isoquintet. In our model three $2^+$ states with isospin 2 are allowed, just above the $0^+$ state.

\emph{In summary:} The model describes the energy spectra of nuclei of mass number 12 especially well.
The rotational band of carbon-12 is very well reproduced, along with some of the
experimentally observed isospin 1 triplets and isospin 2 quintets. However the
observed isotriplets with $J^\pi =0^+$ and $1^-$ do not appear as quantum states
of our $D_{3h}$-symmetric Skyrmion. Neither does the Hoyle state.
They may arise from the quantization of further modes, or appear as quantum states of an alternative Skyrmion.
The molecular rotational band of beryllium-12 in the range 10--20\,MeV above the beryllium-12 
ground state \cite{vonoert} may also be explained in terms of an alternative Skyrmion.

\begin{figure}[h!]
\begin{center}
\includegraphics[width=15.5cm]{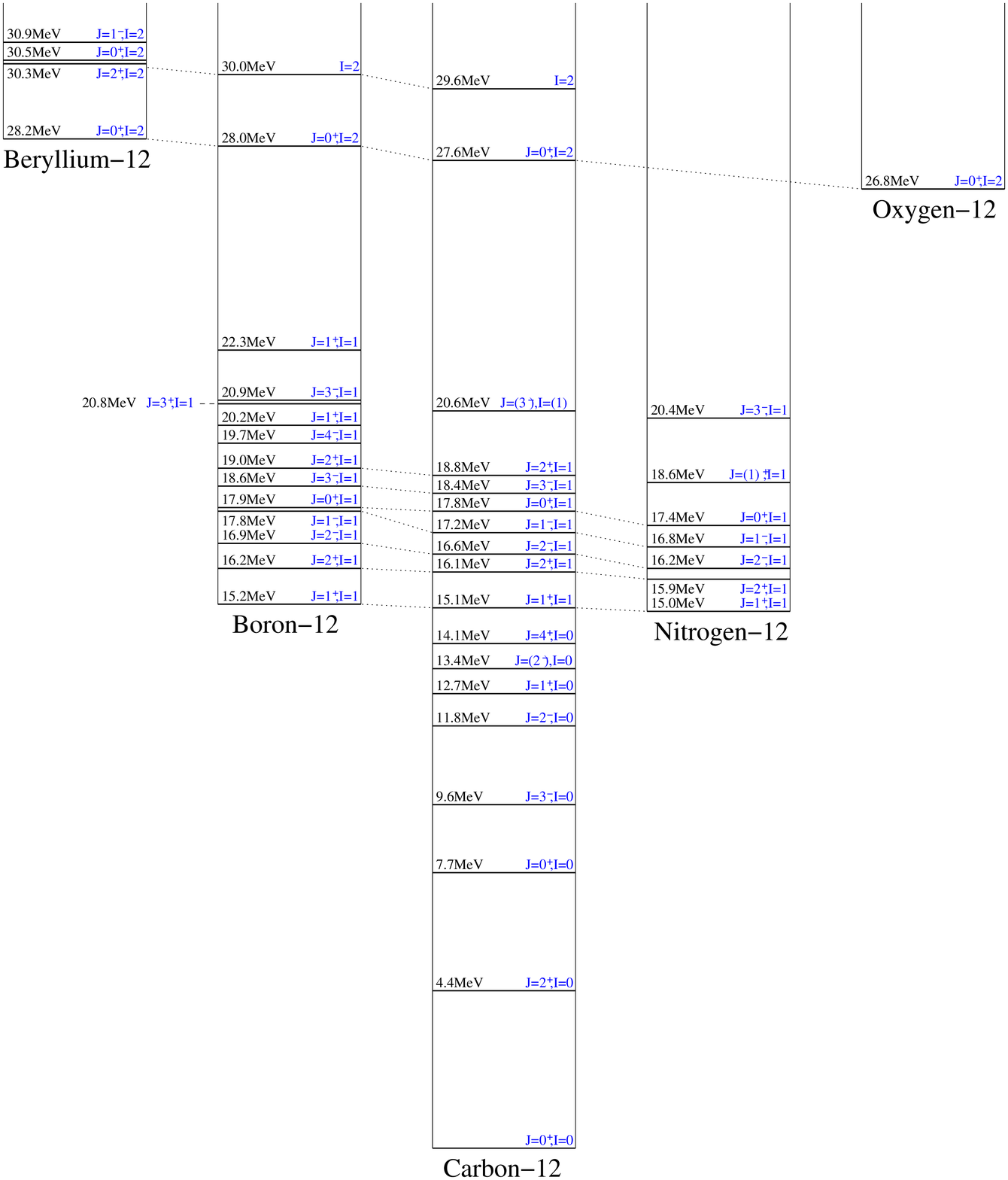}
\caption{Energy level diagram for nuclei of mass number 12.
Mass splittings between nuclei are adjusted to 
eliminate the proton/neutron mass difference and remove Coulomb
effects, as described in Ref. \cite{ajzenberg}.}
\end{center}
\end{figure}

\section{Conclusion}
The Skyrme model provides a model of nuclei which unifies spin and
isospin as collective excitations of topological solitons. We have
been able to reproduce the energy spectra of a number of light nuclei 
to a good degree of accuracy, by quantizing Skyrmions as rigid bodies
in space and isospace, and by parametrising the model separately 
for each mass number. The model is fitted to nuclear charge radii and 
masses. Remarkably, this leads to good predictions for moments of 
inertia, spin splittings and isobar splittings. The quantum states we 
have calculated agree well with what is experimentally observed, with 
the correct spins and parities. Among the successes of the model is
the prediction of the existence and the excitation energies of the 
rotational bands of beryllium-8 and carbon-12. We predict that 
neither beryllium-10 nor carbon-10 has a $1^+$ state, in agreement with experiment.
Our predictions of the quantum states of the nuclear isobars with 
non-zero isospin is quantitatively good, and the spin splittings are
quite good. However, the spin states do not always appear in the 
correct order. A few, so far unobserved states have been predicted,
notably some negative parity states of lithium-8 and boron-8, and some
high spin states including a $4^+$ state of helium-4 and a $6^+$ state
of carbon-12. Several observed states are not predicted by the
rigid-body quantization of Skyrmions. To understand these one will
need to consider further Skyrmion degrees of freedom. 

One can be pleased with the general trend of isospin 
excitations. In each case, the $I=0,1$ splitting is of the order of 
10\,MeV, the $I=0,2$ splitting of the order of 20--30\,MeV,
and the $I=0,3$ splitting of the order of 60\,MeV. 
Isospin splittings for spherically averaged Skyrmions, over a wider range of baryon 
numbers and isospins, have been estimated in Ref. \cite{kopmass}.
The Skyrme model predicts 
that the isobar splittings rise in proportion to $I(I+1)$.
Curiously the experimental data do not rise quite as fast, but 
our prediction for the coefficient of $I(I+1)$ is roughly correct.

It would have been possible to have chosen just one parameter set, as 
in previous work \cite{mmw}. We could have chosen $m$ close to 1, the 
energy scale close to 6\,MeV, and the length scale close to 1.6\,fm, 
for example. This would have provided good qualitative results but 
would not have given such a good fit to charge radii and energy
splittings.

Further work is needed on the electromagnetic form factors 
and transition amplitudes of light 
nuclei within the Skyrme model, as these provide more information 
about the internal structure of nuclei. It would also be desirable to 
consider the effect of vibrational modes and the 
break-up of Skyrmions into clusters, for example, modelling the break-up of 
lithium-6 into helium-4 and the deuteron. Finally, we would like to extend 
this analysis to Skyrmions and nuclei with $B=7,\,9,\,11$ and beyond 12.

\section*{Acknowledgments}
NSM and PMS thank STFC for support under rolling grants ST/G000581/1 and ST/G000433/1.
SWW would also like to thank STFC for financial support.
The parallel computations were performed on COSMOS at the National Cosmology Supercomputing Centre in Cambridge. 

\pagebreak
\begin{appendix}
\section{Numerical Results}
The static energies, mean charge radii and the non-zero elements of the inertia tensors 
for $B=4$, 6, 8, 10 and 12 have been calculated for $m=0.5$, 1
and 1.5, and are shown below. The interpolated values for the preferred values of $m$ are shown in the final columns.
An estimate of the numerical errors can be made by 
calculating the off-diagonal inertia tensor elements, which should be
identically zero in each case. The calculated ratio of off-diagonal 
elements to non-zero diagonal elements is of the order of 
$10^{-2}$ or less.
\subsection{$B=4$}
\begin{table}[ht]
\centering
\begin{tabular}{|l|l|l|l||l|}
\hline
  & $m=0.5$ & $m=1$ & $m=1.5$ & $m=0.820$ \\
\hline
$U_{11}$ & 201 & 151 & 124 & 167\\
$U_{33}$ & 241 & 180 & 146 & 198\\
$V_{33}$ & 928 & 701 & 576 & 771\\
\hline
$\left\langle r^2 \right\rangle ^{1/2}$ & 1.679 & 1.360 & 1.185 & 1.458\\
${\cal{M}}_4$ & 569 & 624 & 681 & 604\\
\hline
\end{tabular}
\end{table} 

\subsection{$B=6$}
\begin{table}[ht]
\centering
\begin{tabular}{|l|l|l|l||l|}
\hline
  & $m=0.5$ & $m=1$ & $m=1.5$ & $m=1.153$ \\
\hline
$U_{11}$ & 305 & 228 & 186 & 211\\
$U_{33}$ & 329 & 245 & 199 & 227\\
$V_{11}$ & 2195 & 1658 & 1362 & 1542\\
$V_{33}$ & 1927 & 1451 & 1190 & 1349\\
$W_{33}$ & $-105$ & $-84$ & $-71$ & $-79$\\
\hline
$\left\langle r^2 \right\rangle ^{1/2}$ & 1.948 & 1.620 & 1.430 & 1.547 \\
${\cal{M}}_6$ & 858 & 946 & 1036 & 973\\
\hline
\end{tabular}
\end{table} 

\subsection{$B=8$}
\begin{table}[ht]
\centering
\begin{tabular}{|l|l|l|l||l|}
\hline
  & $m=0.5$ & $m=1$ & $m=1.5$ & $m=0.832$\\
\hline
$U_{11}$ & 403 & 299 & 243 & 329\\
$U_{22}$ & 374 & 291 & 242 & 315\\
$U_{33}$ & 418 & 326 & 271 & 353\\
$V_{11}$ & 4740 & 4052 & 3490 & 4269\\
$V_{33}$ & 1990 & 1390 & 1109 & 1556\\
\hline
$\left\langle r^2 \right\rangle ^{1/2}$ & 2.316 & 2.017 & 1.787 & 2.109\\
${\cal{M}}_8$ & 1106 & 1213 & 1323 & 1177\\
\hline
\end{tabular}
\end{table} 
\pagebreak
\subsection{$B=10$}
\begin{table}[ht]
\centering
\begin{tabular}{|l|l|l|l||l|}
\hline
  & $m=0.5$ & $m=1$ & $m=1.5$ & $m=0.830$\\
\hline
$U_{11}$ & 511 & 383 & 303 & 421\\
$U_{22}$ & 508 & 380 & 298 & 418\\
$U_{33}$ & 459 & 351 & 285 & 383\\
$V_{11}$ & 4250 & 3120 & 2360 & 3463\\
$V_{22}$ & 5860 & 4520 & 3700 & 4917\\
$V_{33}$ & 5730 & 4400 & 3590 & 4794\\
$W_{33}$ & $-10.4$ & $-4.8$ & 0.7 & $-6.7$\\
\hline
$\left\langle r^2 \right\rangle ^{1/2}$ & 2.455 & 2.047 & 1.745 & 2.174 \\
${\cal{M}}_{10}$ & 1373 & 1516 & 1657 & 1468\\
\hline
\end{tabular}
\end{table}

\subsection{$B=12$}
\begin{table}[h]
\centering
\begin{tabular}{|l|l|l|l||l|}
\hline
  & $m=0.5$ & $m=1$ & $m=1.5$ & $m=0.685$\\
\hline
$U_{11}$ & 588 & 444 & 364 & 527\\
$U_{33}$ & 653 & 500 & 396 & 590\\
$V_{11}$ & 6487 & 5037 & 4087 & 5891\\
$V_{33}$ & 9743 & 7684 & 6240 & 8909\\
$W_{11}$ & $-49$ & $-40$ & $-37$ & $-45$\\
$W_{33}$ & $-42$ & $-35$ & $-40$ & $-38$\\
\hline
$\left\langle r^2 \right\rangle ^{1/2}$ & 2.674 & 2.265 & 1.952 & 2.511\\
${\cal{M}}_{12}$ & 1653 & 1816 & 1982 & 1713\\
\hline
\end{tabular}
\end{table}

\subsection{Quadratic interpolation between three points}
To obtain approximations to the inertia tensors, static energies and mean charge radii for a given value of $m$, we use the method of quadratic interpolation between three points. We know the values of a property $p$ at $m=0.5$, 1 and 1.5, and we make the ansatz
\begin{equation}
p(m) = \alpha_1 + \alpha_2 m + \alpha_3 m^2\,.
\end{equation}
Let ${\mathbf{p}} = (p(0.5),p(1),p(1.5))^{\rm{T}}$. The vector ${\boldsymbol{\alpha}}=(\alpha_1,\alpha_2,\alpha_3)^{\rm{T}}$ is obtained by inverting the expression
\begin{equation}
{\rm{M}}\cdot {\boldsymbol{\alpha}} = {\mathbf{p}}\,,
\end{equation}
where
\begin{equation}
{\rm{M}} = 
\left(
\begin{array}{ccc}
1 & 0.5 & 0.25 \\
1 & 1 & 1 \\
1 & 1.5 & 2.25 
\end{array}
\right)\,,
~~~\hbox{and therefore}~~~
{\rm{M}}^{-1} = 
\left(
\begin{array}{ccc}
3 & -3 & 1 \\
-5 & 8 & -3 \\
2 & -4 & 2 
\end{array}
\right)\,.
\end{equation}
The interpolated values are given in the right-hand columns of the tables in
Appendices A.1 to A.5.

\end{appendix}

\end{document}